\documentclass[aps,prd,twocolumn,nofootinbib,floatfix]{revtex4-2}
\newcommand \color[1]{} % do nothing

\usepackage{graphicx}% Include figure files
\usepackage{amsmath}
\usepackage{amssymb}
\usepackage{amsfonts}

\usepackage{dcolumn}% Align table columns on decimal point
\usepackage{bm}% bold math
\usepackage{bbold} % for mathbb 
\usepackage{float}
\usepackage{pifont}% http://ctan.org/pkg/pifont

 \allowdisplaybreaks  % needs amsmath

\usepackage{hyperref}
\usepackage{scalerel}

\def \VBFONTSIZE {3.5pt}

\newcommand{\A}{ {\scaleto{A}{\VBFONTSIZE}} }
\newcommand{\B}{ {\scaleto{B}{\VBFONTSIZE}} }
\newcommand{\F}{ {\scaleto{F}{\VBFONTSIZE}} }
\newcommand{\M}{ {\scaleto{M}{\VBFONTSIZE}} }
\newcommand{\V}{ {\scaleto{V}{\VBFONTSIZE}} }
\newcommand{\2}{ {\scaleto{2}{\VBFONTSIZE}} }
\newcommand{\ds}{ \displaystyle }

\begin{document}

%\preprint{}

%\title{Stability and scattering of  massive Dirac  bound states in vortices }
\title{Massive Dirac states bound to  vortices by a boson-fermion interaction}


\author{Ethan P. Honda}
\email[]{ehonda@alum.mit.edu}
%\thanks{}
\affiliation{Melbourne, Florida, 32940, USA}
%\affiliation{U.S. Government}


\date{\today}
%\date{August 9, 2025}

\begin{abstract}
Results are presented from numerical simulations of the flat-space nonlinear 
Maxwell-Klein-Gordon-Dirac equations.
The introduction of a boson-fermion interaction allows a scalar vortex to act
as a harmonic trap that can confine massive Dirac bound states.
A parametric analysis is performed to understand the range of 
boson-fermion coupling strengths, Ginzburg-Landau parameters, and fermion effective masses
that support the existence of bound state solutions;
{\color{blue}
results are shown to be comparable to  quasiparticle bound states in gapped Dirac materials.
}
Solutions are time-evolved and are observed to be stable 
until the fermion field $\psi$ becomes large  enough to 
collapse the spontaneously broken vacuum of the condensate.
Head-on scattering simulations are performed, and 
traditional vortex right-angle  scattering is shown to break down with increased 
fermion field strength. For sufficiently large $\psi$ and low velocity, the 
collision of two $m=1$ vortices results  in a pseudostable $m=2$ bound state  
 that eventually becomes unstable and decays back into two $m=1$ vortices.
For large $\psi$ and collision velocity, 
vortex scattering is observed to produce  nontopological (zero winding number) scalar bound states that are 
ejected from the collision.  
The scalar bubbles contain coherent fermion bound states in their interiors  and interpolate between the spontaneously 
broken vacuum of the bulk and the modified vacuum induced by the boson-fermion interaction.
\end{abstract}

%\keywords{}

\maketitle

\section{Introduction}
%%%%%%%%%%%%%%%%%%%%%%%%%%%%%%%%%%%%%%%%%%%%%%%%%%%%%
%%%%%%%%%%%%%%%%%%%%%%%%%%%%%%%%%%%%%%%%%%%%%%%%%%%%%

Vortices are localized configurations with topologically conserved charge that have been widely studied 
across many domains of physics including classical fluid dynamics, condensed matter physics, and particle theory.  
In the context of particle theory,  Nielsen and Olesen explored vortices in the Abelian-Higgs model in an attempt to better understand the 
possible string nature of fundamental particles \cite{Nielsen_197345_Original}.  
% original Abrikosov, abrikosov review.  
%
%
Many authors have since studied the existence, stability, scattering, and other aspects of  
Abelian-Higgs vortices in the context of early universe cosmology, where they 
commonly  appear as cosmic strings in  theories that  undergo spontaneous symmetry breaking 
\cite{
MBHindmarsh_1995,
deVega_PhysRevD.18.2932,
deVega_ClassicalVortexSolution,
Kleidis_ChargedCosmicStrings,
Gleiser_PhysRevD.76.041701,
SHELLARD_1988262,
RUBACK_1988669,
MORIARTY_1988411,
Myers_PhysRevD.45.1355,
Dziarmaga_PhysRevD.49.5609}.
%
%  original N&O, 
%
%
While still unobserved experimentally, cosmic strings are  believed by many to have been present in the early universe,
where they may have influenced the observed  large-scale structure of the universe
\cite{Abbott_PhysRevD.97.102002,Helfer_PhysRevD.99.104028,Pillado_PhysRevD.100.023535}.
Authors have also considered  cosmic strings (vortices) that support the existence of bound fermion states.  % under certain conditions.  
The pioneering work of Nohl explored massive fermion bound states on Abelian vortex lines \cite{Nohl_PhysRevD.12.1840}.
Jackiw and Rossi \cite{Jackiw_1981681_ZeroModes}  explored similar solutions  with 
an interaction that was dimensionally similar to a Yukawa potential but  combined charge-conjugate states
and inspired volumes of work on Majorana  zero modes (MZMs) in both particle theory and condensed matter physics 
\cite{Kleidis_ChargedCosmicStrings,Lozano_PhysRevD.38.601}.
In the standard model, where string configurations
are not inherently topologically stable and must be stabilized dynamically, Weigel et al. realized that the addition of an %Yukawa-like 
 interaction term between  heavy fermions and the boson field actually helps stabilize the string configuration \cite{Weigel_PhysRevLett.106.101601}.
%. 
%

%
While physically different models, when reduced to dimensionless variables the model describing Abelian-Higgs vortices of particle theory 
is the same as  the Ginzburg-Landau-Maxwell model of condensed matter physics,
where vortices  manifest themselves as magnetic flux tubes within superconductors.
These vortices were predicted by Abrikosov in 1957 \cite{Abrikosov_OrigVortex,Abrikosov_RevModPhys.76.975} and experimentally 
verified by Cribier et al. by means of neutron diffraction in 1964 \cite{CRIBIER1964106}.
This groundbreaking work  spawned decades of research and continues to inspire work across many theoretical 
and experimental subfields of  condensed matter physics. 
Of particular  relevance to this work are the advances in Bose-Fermi mixtures and Dirac materials.  
There is a large body of research dedicated to studying Bose-Fermi mixtures where one uses a Bose-Fermi-Hubbard lattice
model that can explain nano-scale interactions between the condensed boson field 
and interacting fermion fields \cite{Roth_PhysRevA.69.021601,Krasnov_BFHModelHilbertSpace,Albus_PhysRevA.68.023606,Lewenstein_PhysRevLett.92.050401,Illuminati_PhysRevLett.93.090406}.
Frequently, a mean field or semi-classical field theory approach is employed when one is interested in %larger (micro-) scale 
effects 
%in the material 
that are large compared to the underlying lattice \cite{Bukov_PhysRevB.89.094502,Chott_PhysRevA.76.010101,Lv_PhysRevA.90.034101}.
Bose-Fermi mixtures describe a wide range of phenomena including charge and spin density waves and
quasiparticles like polarons, excitons, and polaritons \cite{Milczewski_PhysRevA.105.013317}, 
and they are experimentally explored by many in cold-atomic physics using optical traps
\cite{Greiner_OpticalLattice1,Greiner2_PhysRevLett.87.160405,Fehrmann_200423,Cramer_PhysRevLett.93.190405,Albus_PhysRevA.68.023606,Lewenstein_PhysRevLett.92.050401}.
While there is much in the  literature that describes MZMs and other massless Dirac states, there is also a rapidly growing interest in  
massive Dirac fermions as well, particularly in the context of tunable Dirac materials 
\cite{Wehling_02012014,Classen_PhysRevB.93.125119,Ye_MassiveDiracKagome,Yang_MassiveDirac,Lin_PhysRevB.102.155103}. 

The  model being investigated in this work builds upon the  Ginzburg-Landau-Maxwell  and Abelian-Higgs
models %that have been explored thoroughly in the contexts of condensed matter physics and particle theory 
by including a massive Dirac field that repulsively interacts with the scalar field. 
Stable massive Dirac bound state solutions are shown to exist, and a brief look at their scattering properties is provided.
The remainder of this paper is organized as follows.
In Sec. \ref{sec:Formalism} the formalism is defined and the fully general covariant equations of motion 
are presented. 
In Sec. \ref{sec:Stationary} time independent stationary cylindrically symmetric solutions are obtained, approximations are used
to obtain  closed-form solutions, the full solutions are obtained numerically,  and time dependent equations are used 
to determine the stability of the stationary solutions.
In Sec. \ref{sec:Scattering} (2+1) equations are used to perform scattering simulations by boosting the stationary charged vortex 
solutions at each other.  New  bubble-type solutions are observed and briefly discussed.
In an attempt to balance clarity with detail, the  equations of motion are included in the 
main body, while descriptions of numerical methods  appear 
in Appendix \ref{app:NumericalMethods}.
Insight into the self-repulsion and dispersive properties of the one-dimensional (1D)
 free and interacting Dirac fields 
 is provided in  Appendix \ref{app:SelfInteraction}  
 to aid in understanding results from the model being explored in this work.

%%%%%%%%%%%%%%%%%%%%%%%%%%%%%%%%%%%%%%%%%%%%%%%%%%%%%
%%%%%%%%%%%%%%%%%%%%%%%%%%%%%%%%%%%%%%%%%%%%%%%%%%%%%
\section{General formalism, definitions, and conventions \label{sec:Formalism}}
%%%%%%%%%%%%%%%%%%%%%%%%%%%%%%%%%%%%%%%%%%%%%%%%%%%%%
%%%%%%%%%%%%%%%%%%%%%%%%%%%%%%%%%%%%%%%%%%%%%%%%%%%%%

The  model being studied  is constructed by  
 minimally coupling the
 Maxwell, 
 spontaneously broken Ginzburg-Landau,  
 massive  Dirac, 
 and boson-fermion interaction Lagrangians,
\begin{eqnarray}
\mathcal{L} &=& 
\sqrt{-g}\left( L_{M} +
L_{GL} +
L_{D} +
L_{GLD} 
\right),
\label{eqn:OverallLagrangian}
\end{eqnarray}
where 
\begin{eqnarray}
%%%%%%%%%%%%%%
%%%%%%%%%%%%%%
{L}_{M} 
&=&
-\frac{\epsilon_0 c^2}{4} F^{\mu\nu}F_{\mu\nu},
 \\
%%%%%%%%%%%%%%
{L}_{GL} 
&=&
-\frac{\hbar^2}{2m_\B}g^{\mu\nu} D_\mu\phi\left(D_\nu\phi\right)^* + \alpha_\B\phi^2 - \frac{\beta_\B}{2}\phi^4,
 \\
%%%%%%%%%%%%%%
{L}_{D} 
&=& 
\frac{ic\hbar}{2}\left[
 \bar{\psi} \gamma^\mu D_\mu\psi  - 
 \left(D^*_\mu \bar{\psi} \right) \gamma^\mu \psi 
\right] 
%\nonumber \\ &&
-   m_\F c^2\bar{\psi} \psi, \text{ and}
\nonumber \\
 \\
%%%%%%%%%%%%%%
{L}_{GLD} 
&=& 
 -\frac{1}{2}\mu_i (\phi^*\phi) (\bar{\psi}\psi);
\end{eqnarray}
the electromagnetic field strength tensor, scalar gauge covariant derivative, and fermion gauge covariant derivative
are defined to be
\begin{eqnarray}
%%%%%%%%%%%%%%%%
F_{\mu\nu} &=& \partial_\mu A_\nu - \partial_\nu A_\mu,
\\
%%%%%%%%%%%%%%%%
D_\mu\phi &=&  \partial_\mu\phi  - \frac{iq_2}{\hbar} A_\mu\phi, \text{ and}
\\ 
%%%%%%%%%%%%%%%%
D_\mu\psi &=&  \partial_\mu \psi - \frac{iq_1}{\hbar}A_\mu \psi + \Gamma_\mu\psi;
%%%%%%%%%%%%%%%%
\end{eqnarray}
and the spinor affine connection, spin connection, gamma matrices in a coordinate basis, and spinor adjoint are given by
\begin{eqnarray}
%%%%%%%%%%%%%%%%
\Gamma_\mu   &=& -\frac{1}{8} \omega_{\mu\A\B}   \left[\tilde{\gamma}^A, \tilde{\gamma}^B\right],  \\
%%%%%%%%%%%%%%%%
 \omega_{\mu\A\B}  &=&
g_{\nu\alpha} {e^\alpha}_{A}  \left( \partial_\mu{e^\nu}_B  + {\Gamma^\nu}_{\mu\lambda} {e^\lambda}_B  \right), \\
%%%%%%%%%%%%%%%%
 %&=&
%%%%%%%%%%%%%%%%
\gamma^\alpha &=& ({e^\alpha}_A) \tilde{\gamma}^A, \text{ and} \label{eqn:GammaConversion}\\
%\tilde{\gamma}^A &=& ({e_\alpha}^A) \gamma^\alpha \\
\bar{\psi} &=& \psi^\dag \gamma^0.
%%%%%%%%%%%%%%%%
\end{eqnarray}
For full generality, the following dimensionless variables are defined,
\begin{eqnarray}
\hat{x}^\mu	&=& \left( \frac{1}{\xi} \right)x^\mu,  \label{eqn:DimionsionlessCoords} \\
\hat{A}_\mu	&=& \left( \frac{q_2\Lambda}{\hbar}\right) A_\mu,  \\
\hat{\phi}		&=& \left( \frac{\beta_\B}{\alpha_\B}\right)^{1/2} \phi, \text{ and}\\
\hat{\psi}		%&=& \frac{\kappa_{d1}}{\sqrt{2}}  \left( \frac{ \beta_B  }{  \alpha_B}\right)^{1/2}\psi  \\
			&=&\left( \frac{ c \xi }{\hbar } \right)  \left( \frac{m_\B m_\F \beta_\B  }{  \alpha_\B}\right)^{1/2}\psi,  
			\label{eqn:PsiDimensionless}
\end{eqnarray}
and  physical parameters of condensate number density, 
London penetration depth, and coherence length, are defined to be
%respectively,
%
\begin{eqnarray}
\phi_0^2	&=& \frac{\alpha_\B}{\beta_\B},  
\label{eqn:phi2alphabeta} \\
%\Lambda^{-2} 	&=& \frac{q_2^2 \phi_0^2 }{m_B \epsilon_0 c^2} \\
\Lambda 	&=& \left( \frac{m_\B \epsilon_0 c^2}{q_2^2 \phi_0^2 } \right)^{1/2}, \text{ and}
\\
%\xi^2			&=& \frac{\hbar^2}{2 m_B \alpha_B} \label{eqn:xi2alpha} \\
\xi			&=& \left( \frac{\hbar^2}{2 m_\B \alpha_\B} \right)^{1/2}, \label{eqn:xi2alpha}
\end{eqnarray}
which gives rise to the dimensionless model parameters
\begin{eqnarray}
\kappa &=& \frac{\Lambda}{\xi}, \\
\kappa_{d} &=&  \left( \frac{\hbar}{m_\F c \xi  }\right) 
= \frac{1}{2\pi} \left(\frac{\lambda_C}{\xi}\right)  
=   \frac{\lambdabar_C}{\xi}, \text{ and}
\label{eqn:kappad}
\\
\kappa_m &=& \frac{\mu_i \phi_0^2}{m_\F c^2}, 
\end{eqnarray}
where $\kappa$ is the traditional Ginzburg-Landau (GL) parameter, 
%$\kappa_d^{-1}$ is an effective fermion mass,  
$\kappa_d$ is the scaled Compton length of the fermion quasiparticle,  
and
$\kappa_m$ measures the boson-fermion coupling strength.
In practice it is also helpful to use $\kappa_d^{-1}$, which functions as an 
effective dimensionless fermion mass.
From these expressions, one can obtain the  dimensionless actions
\begin{eqnarray}
%%%%%%%%%%%%%%%%%%%%%%
\hat{{L}}_{M} 
&=&
-\frac{1}{4}
\hat{F}^{\mu\nu} \hat{F}_{\mu\nu},
\label{eqn:Maxwell_Lagrangian}
 \\
%%%%%%%%%
\hat{{L}}_{GL} 
 &=&
-\frac{1}{2}  
g^{\mu\nu} \hat{D}_\mu\hat{\phi}\left(\hat{D}_\nu\hat{\phi}\right)^* 
+ \frac{1}{2}\hat{\phi}^2 
- \frac{1}{4} \hat{\phi}^4,
\label{eqn:GL_Lagrangian}
  \\
%%%%%%%%
%%%%%%%%%%%%%%
\hat{\mathcal{L}}_{D} 
&=& 
 \frac{i}{2}\kappa_{d}
\left[
 \hat{\bar{\psi}} \gamma^\mu  \hat{D}_\mu\hat{\psi} 
 -  \left(\hat{D}^*_\mu \hat{\bar{\psi}} \right) \gamma^\mu \hat{\psi} 
  \right] 
-  
\hat{\bar{\psi}} \hat{\psi},\text{ and} \hspace{5mm}
\label{eqn:Dirac_Lagrangian}
\\
%%%%%%%%%%%%%%
\hat{\mathcal{L}}_{GLD} 
&=& 
 -\frac{1}{2} \kappa_m  \left(\hat{\phi}^*\hat{\phi}\right)  \left(\hat{\bar{\psi}}\hat{\psi}\right),
\label{eqn:GLD_Lagrangian}
%\\
%%%%%%%%%%%%%%%
\end{eqnarray}
%
%where it has been assumed that $q_2 = 2 q_1$. 
{\color{blue}
where  to preserve local $U(1)$ gauge invariance it is assumed that $q_2 = 2 q_1$. 
While many normalizations of the Dirac field may be appropriate, 
the approach assumed in this work is  
\begin{eqnarray}
\int_\V d^3x \left( \psi^\dag\psi\right) &=& 1,
\label{eqn:DimensionfulNorm}
\end{eqnarray}
where $\psi$ is the {dimensionful}  field ($[\psi]=L^{-3/2}$)
representing a single fermionic quasiparticle. 
Using (\ref{eqn:PsiDimensionless}) and (\ref{eqn:phi2alphabeta}) one gets
\begin{eqnarray}
%%%%%%%%
\int_{V} d^3\hat{x} \left( \hat{\psi}^\dag \hat{\psi}\right) &=&  
\left(m_\F \right)
%m_\F 
\left(\frac{m_\B}{  \phi_0^2\xi}  \right)
\left(\frac{c^2}{ \hbar^2} \right),
\label{eqn:DimensionlessNormalization}
%=
%\text{(conserved over time)}
%\nonumber 
\end{eqnarray}
which will be conserved over time by the equations of motion.
Throughout this work the amplitude of $\hat{\psi}$ will be parametrically 
varied but it should be noted that changing the norm of the dimensionless 
field does not affect (\ref{eqn:DimensionfulNorm}); changing the norm 
of  $\hat{\psi}$ simply represents a change of the underlying  model parameters.
For example, one might set the  properties of the bulk ($m_\B$, $\xi$, and $\phi_0^2$) to match a material of interest
and  use  different values of $\hat{\psi}$ as a means to explore what  effective masses ($m_\F$)
 support  fermion bound state solutions.
}

Moving forward with dimensionless variables,  the  $\hat{}$\,s will be omitted  for 
clarity.  
The covariant equations of motion
for the complex scalar field are given by
\begin{eqnarray}
%%%%%%%%%
%\frac{1}{\sqrt{-g}}\partial_\mu\left( \sqrt{-g}g^{\mu\nu}\partial_\nu\phi\right)
\Box\phi  
&=& 
 2i \kappa^{-1} A^\rho   \partial_\rho  \phi
+  i \kappa^{-1} \phi  \nabla_\rho A^\rho  
+ \kappa^{-2} \phi A_\mu A^\mu 
\nonumber \\ &&
- \left(1-\kappa_m\hat{\psi}^2 \right) \phi + (\phi^*\phi)\phi \text{ \  and}
\\
%%%%%%%%%
%\frac{1}{\sqrt{-g}}\partial_\mu\left( \sqrt{-g}g^{\mu\nu}\partial_\nu\phi^*\right)
\Box\phi^*  
&=& 
- 2i\kappa^{-1}  A^\mu \partial_\mu\phi^* 
-  i \kappa^{-1}  \phi^*  \nabla_\rho A^\rho  
+ \kappa^{-2} \phi^* A_\mu A^\mu 
\nonumber \\ &&
-\left(1-\kappa_m\hat{\psi}^2 \right) \phi^* +  (\phi^*\phi)\phi^*,
%\nonumber \\
\end{eqnarray}
where
\begin{eqnarray}
\Box\phi_i  &=& 
\frac{1}{\sqrt{-g}}\partial_\mu\left( \sqrt{-g}g^{\mu\nu}\partial_\nu\phi\right).
\end{eqnarray}
The covariant Maxwell equations are given by
\begin{eqnarray}
%%%%%%%%%%%%%
\nabla_\alpha F^{\beta\alpha} 
&=& 
 -\frac{i}{2\kappa}  g^{\beta\mu}  \left( \phi^*\partial_\mu \phi- \phi \partial_\mu \phi^*   \right)
-   \kappa^{-2} A^\beta \phi^*\phi 
\nonumber \\ && 
+ \frac{1}{2}\kappa_{d} \kappa^{-1}
\left( \bar{\psi}\gamma^\beta \psi  \right) \text{ and}
\\
%%%%%%%%%%%%
\partial_{[\alpha}  F_{\mu\nu]} &=& 0,
\end{eqnarray}
and the generally covariant Dirac equation is given by
\begin{eqnarray}
 i  \gamma^\mu \left(  \partial_\mu  -  \frac{i}{2\kappa} A_\mu  +  \Gamma_\mu   \right) \psi-   
\kappa_{d}^{-1} \left(1+\frac{\kappa_m\hat{\phi}^2}{2} \right)\psi   &=& 0.  
\nonumber \\ 
%\\
\end{eqnarray}
All spinors and gamma matrices are presented  in the Dirac representation.  The gamma matrices in
an orthonormal basis are given by
\begin{equation}
%%%%%%%%%
\tilde{\gamma}^0  = \left(
\begin{array}{cc}
\mathbb{1}_\2 & \mathbb{0}_\2 \\
\mathbb{0}_\2 & -\mathbb{1}_\2 
\end{array}
\right),
%%%%%%%%%
\hspace{2mm}
%%%%%%%%%
\tilde{\gamma}^k  = \left(
\begin{array}{cc}
\mathbb{0}_\2 & \mathbb{\sigma}^k \\
-\mathbb{\sigma}^k & \mathbb{0}_\2 
\end{array}
\right),
%%%%%%%%%%
\end{equation}
where %$\gamma^5 = i \gamma^0\gamma^1\gamma^2\gamma^3$, and 
\begin{equation}
%%%%%%%%%
\sigma^1  = \left(
\begin{array}{cc}
 0 & 1 \\
1 & 0 
\end{array}
\right),
%%%%%%%%%
\hspace{1mm}
%%%%%%%%%
\sigma^2  = \left(
\begin{array}{cc}
 0 & -i \\
i & 0 
\end{array}
\right),
%%%%%%%%%
\text{  and  }
%%%%%%%%%
\sigma^3  = \left(
\begin{array}{cc}
 1 & 0 \\
0 & -1
\end{array}
\right)
\end{equation}
are converted to coordinate frames as needed using tetrads (\ref{eqn:GammaConversion}).
Section \ref{sec:Stationary} will use Cartesian-aligned tetrads with cylindrical coordinates, 
which allow the spin connection ($\omega_{\mu\A\B}$) and spinor affine connection ($\Gamma_\mu$)
to be zero, but it should  be noted that $\gamma^\mu\neq \tilde{\gamma}^\mu$.
Section \ref{sec:Scattering} will use Cartesian-aligned tetrads with Cartesian coordinates, which gives 
$\omega_{\mu\A\B} = \Gamma_\mu = 0$ and  $\gamma^\mu = \tilde{\gamma}^\mu$.

The spacetime metric is assumed to be flat with negative signature $(-,+,+,+)$, which 
gives 
\begin{eqnarray}
\left\{\gamma^\mu,\gamma^\nu\right\} &=& -2g^{\mu\nu}\mathbb{1}_4
\end{eqnarray}
for the Dirac algebra.
Finally,  only GL parameter values $\kappa \geq 1$ will be considered, which will focus the analysis on 
Type-II superconductor types of behaviors,
and all Maxwell equations are solved using the Lorentz gauge.  

%\lipsum[10]

%%%%%%%%%%%%%%%%%%%%%%%%%%%%%%%%%%%%%%%%%%%%%%%%%%%%%%%%%%%%%
%%%%%%%%%%%%%%%%%%%%%%%%%%%%%%%%%%%%%%%%%%%%%%%%%%%%%%%%%%%%%
\section{Stationary axisymmetric solutions \label{sec:Stationary}}
%%%%%%%%%%%%%%%%%%%%%%%%%%%%%%%%%%%%%%%%%%%%%%%%%%%%%%%%%%%%%
%%%%%%%%%%%%%%%%%%%%%%%%%%%%%%%%%%%%%%%%%%%%%%%%%%%%%%%%%%%%%
%

This section explores the existence and stability of stationary massive Dirac bound states in vortices.
Time independent equations are derived, and closed form approximate solutions and full numerical  solutions
are presented.  
A parametric analysis is performed to understand where bound states exist as a function of boson-fermion 
coupling strength, GL parameter, and effective fermion mass.
Time dependent equations are derived and used to demonstrate stability of solutions and where they 
become unstable.

%%%%%%%%%%%%%%%%%%%%%%%%%%
\subsection{Stationary equations of motion}
%%%%%%%%%%%%%%%%%%%%%%%%%%

A general ansatz for a stationary fermion bound state confined to a 
scalar vortex can be  given by
\begin{eqnarray}
%%%%%%%%%%%%%%%%%%
\phi(R,\theta) &=& \phi(R) e^{im\theta}, \text{ and} 
\\
%%%%%%%%%%%%%%%%%%
\psi(t,R,\theta)  
&=& 
\left(
\begin{array}{c}
\Psi_1(R) e^{im_1\theta} e^{-i \omega_1 t}  \\
\Psi_2(R) e^{im_2\theta} e^{-i \omega_2 t}  \\
\pm i \Psi_3(R) e^{im_3\theta} e^{-i\omega_3 t}  \\
\pm i\Psi_4(R) e^{im_4\theta} e^{-i\omega_4 t} \\
\end{array}
\right).
\label{eqn:SpinorAnsatz}
\end{eqnarray}
The desire to have stable scalar vortices %in the fermion weak-field limit 
leads to selecting a 
vortex winding number of $m=1$.
{\color{blue}
The $\omega_i$ are all set to $\omega$ so that the time-dependent equations become 
separable in time.
}
The conditions 
\begin{eqnarray}
m_4 &=& m_1+1 \text{ and} \\
m_2 &=& m_3+1
\end{eqnarray}
allow for angular separation of the partial differential equations and also lead to the 
fermion solutions being eigenstates of total angular momentum in the $z$-direction,  $\hat{J}_z$.
The addition of $\pm i$ to  $\psi_3$ and $\psi_4$ is simply an overall phase that simplifies the
field equations that follow.
One can then assume  $A_z=0$, which decouples $(\psi_1,\psi_4)$ and $(\psi_2,\psi_3)$, 
which allows one to set $\psi_2=\psi_3=0$.
Finally, setting $m_1=0$ leads to the fermion solutions being $j_z=+\frac{1}{2}$ 
eigenstates.
These assumptions lead to the stationary equations of motion
\begin{eqnarray}
%%%%%%%%%%%%%%%%%%%%%%%%%%
\partial_R\chi &=& 
 -2 \kappa^{-1} \frac{\tilde{A}_\theta}{R}  {\phi}  
+ \kappa^{-2} {\phi}  \left( -A_t^2 +  \tilde{A}_\theta^2 \right)
\nonumber \\ &&
- \left( 1 - \kappa_m \bar{\psi}\psi\right) {\phi} +  {\phi}^3, 
\label{eqn:FirstStataionary}
\\
%%%%%%%%%%%%%%%%%%%%%%%%%%
 \frac{\partial}{\partial R^2} \left( R {\phi} \right) 
 &=& \frac{1}{2}\chi, 
\\
%%%%%%%%%%%%%%%%%%%%%%%%%%
%\\
\partial_RB^z
&=&
-  \frac{ \phi^2 }{\kappa R} 
+ \kappa^{-2}\tilde{A}_\theta  \phi^2 
%\nonumber \\ &&
-\frac{\kappa_{d}}{\kappa} 
\left(\Psi_1 \Psi_4 \right),
\\
%%%%%%%%%%%%%%%%%%%%%%%%%%
\frac{1}{R} \partial_R(R \tilde{A}_\theta ) &=& B_z,\\
%%%%%%%%%%%%%%%
%%
%%%%%%%%%%%%%%%%%%%%%%%%%%
\frac{1}{R} \partial_R\left(  R E^R \right)
&=&
\kappa^{-2}A_t {\phi}^2 
+
\frac{\kappa_{d}}{2\kappa} \left( 
\Psi_1^2  + \Psi_4^2 
\right),
\\
%%%%%%%%%%%%%%%%%%%%%%%%%%
\partial_R A_t  &=& E_R,  \\
%%%%%%%%%%%%%%%%%%%%%%%%%%%%%%%%%%%%%%%%%%%%%%%%%%%%
%%%%%%%%%%%%%%%%%%%%%%%%%%%%%%%%%%%%%%%%%%%%%%%%%%%%
 \partial_R{\Psi}_1
&=&
-\left(  \omega
+ \frac{1}{2\kappa}A_t
+ \kappa_d^{-1}  \left(1+\frac{\kappa_m\hat{\phi}^2 }{2}\right) \right){\Psi}_4
\nonumber \\ &&
- \frac{1}{2\kappa}\tilde{A}_\theta{\Psi}_1, \text{ and}
\label{eqn:StationaryPsi1}
\\
%%%%%%%%%%%%%%%%%%%%%%%%%%%%%%%%%%%%%%%%%%%%%%%%%%%%
\frac{1}{R}\partial_R\left( R {\Psi}_4\right)
%\left(\partial_R   + \frac{m+1}{R} \right)\left( \pm \hat{\Psi}_4\right)   
&=&
\left( 
\omega 
+ \frac{1}{2\kappa}A_t
- \kappa_d^{-1}  \left(1+\frac{\kappa_m\hat{\phi}^2}{2} \right)
\right){\Psi}_1
\nonumber \\ &&
+ \frac{1}{2\kappa}  \tilde{A}_\theta  {\Psi}_4,
\label{eqn:StationaryPsi4}
\label{eqn:LastStataionary}
\end{eqnarray}
where 
\begin{eqnarray}
 \bar{\psi}\psi  &=& {\Psi}_1^2 - {\Psi}_4^2, \text{ and} \\
 \tilde{A}_\theta &=& A_\theta / R = A^\theta R.
\end{eqnarray}
The field $\chi$ was introduced to help with regularity at the origin when 
solving these equations numerically.
While it is  helpful to be able to refer to the fermion bound states ($\psi_1$ and $\psi_4$) 
and the scalar vortex solutions ($\phi$) as distinct entities,  when being discussed as a 
single entity their combination will hereafter be referred to as a ``charged vortex."

%%%%%%%%%%%%%%%%%%%%%%%%%%
\subsection{Approximate closed-form solutions \label{subsec:ClosedForm}}
%%%%%%%%%%%%%%%%%%%%%%%%%%

To get an approximate understanding of when bound states exist, 
an effective potential 
is obtained by taking an additional spatial derivative of (\ref{eqn:StationaryPsi1}) and substituting repeatedly
to put the resulting equation in the form 
\begin{eqnarray}
\frac{\partial^2\Psi_1}{\partial R^2} 
%&=& V_1 \Psi_1 + V_4 \Psi_4 
&\approx& f_1 \Psi_1,
\label{eqn:ApproxWaveEquation}
\end{eqnarray}
where
\begin{eqnarray}
%%%%%%%%%%%%%%%%%%%%%%%%%%%%%%%%%%%%
f_1
 &=&
\kappa_d^{-2}  \left(1+\frac{1}{2} \kappa_m\hat{\phi}^2 \right)^2 
-\left( \omega + \frac{1}{2\kappa}A_t \right)^2 
\nonumber \\ &&
+\frac{\tilde{A}^2_\theta}{4\kappa^2}
-\frac{1}{2\kappa}\left(B_z - \frac{\tilde{A}_\theta}{R}\right),
\label{eqn:EffectivePotential}
\end{eqnarray}
and terms that multiply  $\psi_4$ are neglected
since for most solutions, 
especially for large effective fermion mass, 
the $\Psi_4$ terms are small relative to $\Psi_1$.
Based on the linear properties of the scalar field at the origin of the vortex, the following form for the scalar 
field is a reasonable approximation,
\begin{eqnarray}
\phi &\approx& \frac{R}{a},
\end{eqnarray}
which to leading order in $R$ creates a spatially harmonic potential for a positive (repulsive) boson-fermion interaction.
In the large $\kappa$ limit, the gauge field contributions are small, which leads to
\begin{eqnarray}
f_1
%%%%%%%%%%%%%
&\approx&
\kappa_d^{-2} \left( 
1 + \kappa_m\left( \frac{R}{a}\right)^2   
\right) - \omega^2. 
\label{eqn:F1Approx}
\end{eqnarray}
Equations  (\ref{eqn:ApproxWaveEquation}) and (\ref{eqn:F1Approx}) are known to have Gaussian solutions, 
\begin{eqnarray}
\Psi_1(R) &=& \Psi_{1,0}\exp\left(-\frac{R^2}{2\sigma^2}\right),
\label{eqn:PsiGauHarmonicTrap}
\end{eqnarray}
for real constant $\Psi_{1,0}$ and when subjected to the constraints
\begin{eqnarray}
%%%%%%%%%%%%%%%
\omega^2 &=& \sigma^{-2} + \kappa_d^{-2} \text{ \ and}  \label{eqn:DispersionRelation}  \\
%%%%%%%%%%%%%%%%%%
\sigma^{-4} &=& \frac{\kappa_m\kappa_d^{-2}}{a^2}.
 \label{eqn:TrapR2Relation} 
 \end{eqnarray}
{\color{blue}
Equation  (\ref{eqn:DispersionRelation}) is the dispersion relation and corresponds to 
the  energy-momentum  relation %$E(p)$ 
one might use in condensed matter systems, $E(p)$,
where $\omega \propto E$, \hbox{$\sigma^{-2} %\propto k^2 
\propto p^2$},  $(\kappa_d^{-1})^2 \propto m_\F^2$,
and  $\kappa_d^{-1}>0$ describes a gapped material.
}

Knowing that bound states can exist when $f_1$ is negative within the trap and positive outside
the trap, using $f_1(0)<0$ and $f_1(R_{\rm max})>0$ allows one to obtain a condition on 
 $\omega$ that should support fermion bound states,
\begin{eqnarray}
\omega^2 &>& \kappa_d^{-2} \text{ \ and} \label{eqn:Omega2GreaterThan} \\
\omega^2 &<& \kappa_d^{-2}\left(  1+\frac{1}{2}\kappa_m\right)^2, \label{eqn:Omega2LessThan}
\end{eqnarray}
or more simply,
\begin{equation}
\kappa_d^{-1} <  \omega  <  \kappa_d^{-1}\left(  1+\frac{1}{2}\kappa_m\right). \label{eqn:MassBand}
\end{equation}
Solving  (\ref{eqn:Omega2LessThan}) for the roots of $\kappa_m$, one gets
\begin{eqnarray}
%%%%%%%%%%%%%%
\kappa_m^+ &\geq&  2 \delta_\omega \text{ and} \label{eqn:MinConfinement}\\
\kappa_m^- &\leq&  -4 - 2 \delta_\omega, 
\end{eqnarray}
where one can define
\begin{eqnarray}
\delta_\omega &=&  \frac{\omega}{\kappa_d^{-1}} -1.
\end{eqnarray}
%a positive constant.  
The focus is on the positive (repulsive) root for fermion states so they can be  
bound within the vortex, and it can be seen that there are no bound states when $\kappa_m=0$.
{\color{blue}
Equation (\ref{eqn:MassBand}) describes the  energy band of allowable bound states and is bounded 
from below by the  effective mass  ($\kappa_d^{-1}\propto m_\F$)
and bounded from above by the effective mass increased 
by a term proportional to the boson-fermion interaction strength ($\kappa_m$).
The size of the energy band goes to zero in the limit that the interaction strength goes to zero ($\kappa_m\rightarrow 0$),
implying a lack of bound states when there is no boson-fermion interaction.
}

Another aspect of these solutions that can be understood in closed form is the impact of the
fermion field on the spontaneously broken scalar vacuum.  
The shape of the scalar field vortex and the location of its vacuum expectation value is determined by the potential
\begin{eqnarray}
V(\phi) &=& - \frac{1}{2}\left( 1- \kappa_m\bar{\psi}\psi\right)\phi^2 + \frac{1}{4}\phi^4,
\label{eqn:VphiSSB}
\end{eqnarray}
where for small $\kappa_m\bar{\psi}\psi$ the expectation value approaches the %well known 
uncharged vortex solution with expectation value  $\phi_0=1$.
For larger fermion field strength  %number densities, 
the shape of the potential will change, eventually
reaching a point where the spontaneously broken vacuum collapses back to $\phi_0=0$
as  $\kappa_m\bar{\psi}\psi\rightarrow 1$.  
This effectively destroys superconductivity in a larger region than just the center of the vortex, 
 flattening out the harmonic trap for the fermion field, and suggests a loss of containment 
 of the fermion field by the trap.
Using (\ref{eqn:TrapR2Relation}) the condition on the fermion field
for when this occurs
becomes
\begin{eqnarray}
\psi_{1,0}^2 &=&  \frac{\sigma^{4} \chi_0^2}{4} (\kappa_d^{-1})^2, %\approx (\kappa_d^{-1})^2 
\label{eqn:FermionUpperBound}
\end{eqnarray}
where $\chi_0$ is the value of $\chi$ at the origin and
$\sigma^{4} \chi_0^2/4$ is close to unity when the width of the fermion bound state 
is close to the width of the vortex. 
{\color{blue}
Returning briefly to dimensionful variables,  %Equations
(\ref{eqn:kappad}),
(\ref{eqn:DimensionlessNormalization}), and
(\ref{eqn:FermionUpperBound})
imply that the range of effective masses  that support the existence of bound states has
an upper limit determined by the parameters of the bulk,
\begin{eqnarray}
0 \leq m_\F \lesssim  \left( \frac{m_\B}{\phi_0^2 \xi^3}\right).
\end{eqnarray}
For typical bulk parameters $\phi_0^2 \approx 10^{30} {\rm m}^{-3}$, $\xi \approx 10^{-9}{\rm m}$,
and   $m_\B=2m_e$,  this condition becomes 
\begin{eqnarray}
0 \leq m_\F \lesssim  2\times 10^{-3} m_e,
\end{eqnarray}
which is comparable  to the effective masses of quasiparticles in many gapped Dirac materials 
\cite{KUMAR_2014S136,Fischer_ArmchairGraphene}.
In summary,  for the approximate bound states discussed here,
the effective mass $\kappa_d^{-1}\propto m_\F$ sets the  mass gap while
the interaction strength $\kappa_m$ determines  the energy band of allowed bound states.
}

%%%%%%%%%%%%%%%%%%%%%%%%%%
\subsection{Stationary numerical solutions \label{subsec:Stationary}}
%%%%%%%%%%%%%%%%%%%%%%%%%%

%%%%%%%%%%%%%%%%%%%%%%%%
\begin{figure}[]
\begin{center}
\includegraphics[width=80mm]{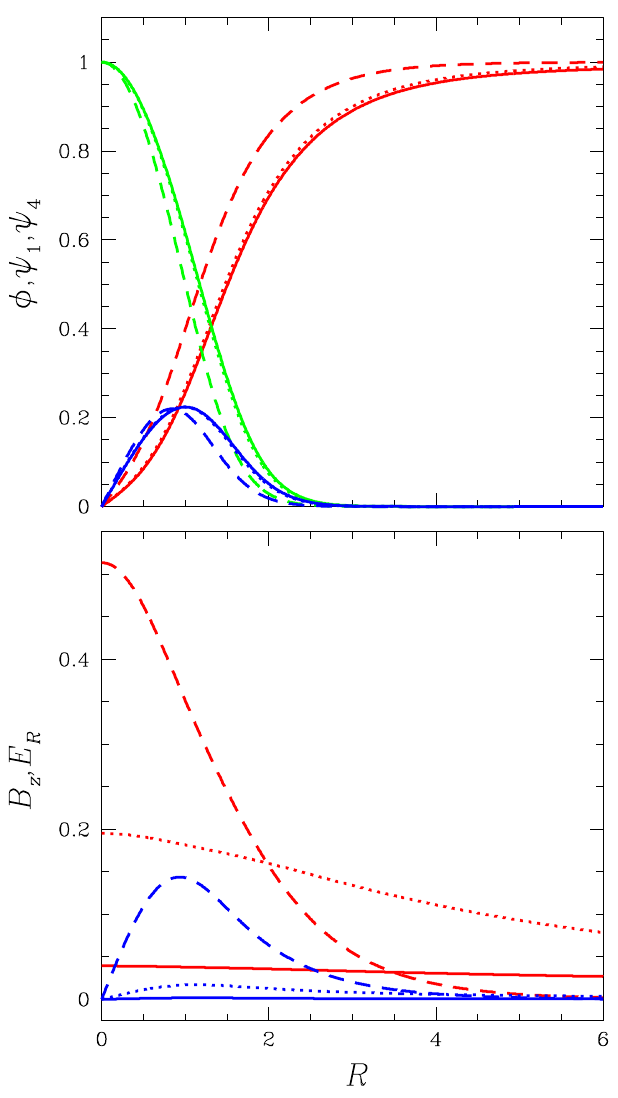}
\caption{
Plots of scalar field $\phi$ (top, red), fermion field components $\psi_1$ (top, green), 
 $\psi_4$ (top, blue), electric field $E_R$ (bottom, red), and magnetic field $B_z$ (bottom, blue) 
for a
bound state solution with effective mass $\kappa_d^{-1}=1$, and GL parameter 
values $\kappa=1,10,100$ in dashed, dotted, and solid lines, respectively.  The confinement strength,
$\kappa_m$, is chosen to give a fermion radius of unity for $\kappa=1$.
 } 
\label{fig:VortNoPsi.pdf}
\end{center}
\end{figure}
%%%%%%%%%%%%%%%%%%%%%%%%

This section %subsection 
presents the  numerical solutions to the full stationary 
equations of motion, demonstrating the existence of bound state solutions
across a wide range of GL parameters ($\kappa$), effective fermion masses ($\kappa_d^{-1}$), 
and boson-fermion interaction strengths ($\kappa_m$).
Starting with a single point in this large parameter space, 
Figure \ref{fig:VortNoPsi.pdf}  displays the basic attributes of 
typical  fermion bound states confined to the core of a scalar vortex.
The scalar field, fermion field components, and electric and magnetic fields are shown 
 for a range of GL parameters, $\kappa=1,10,100$ %
and effective fermion mass $\kappa_d^{-1}=1$.  
The confinement strength $\kappa_m$ is tuned to obtain a  unit radius at half-max for the $\psi_1$ component of 
the fermion field for $\kappa=1$.
The scaling of the spatial coordinates by the coherence length  (\ref{eqn:DimionsionlessCoords})  ensures that the vortex core
will remain roughly on the order of unity.  However, as one increases the GL parameter, the  London penetration depth increases,
which increases the penetration of the field into the bulk while keeping the topologically conserved magnetic flux constant.
This reduces the field strength within the vortex core so that
in the large-$\kappa$ limit, the shape of the at-rest bound state  fermion becomes independent 
of the electromagnetic fields.
Hereafter, the radius of  $|\psi_1(R)|$ at half-max is denoted  $R_\psi$.

%

%%%%%%%%%%%%%%%%%%%%%%%%%%%%%%%%%%%%%%%%%
\begin{figure}[]
% source location: ~/cpp/cv1dx/figs/Rd_vs_kdm1
\begin{center}
\includegraphics[width=80mm]{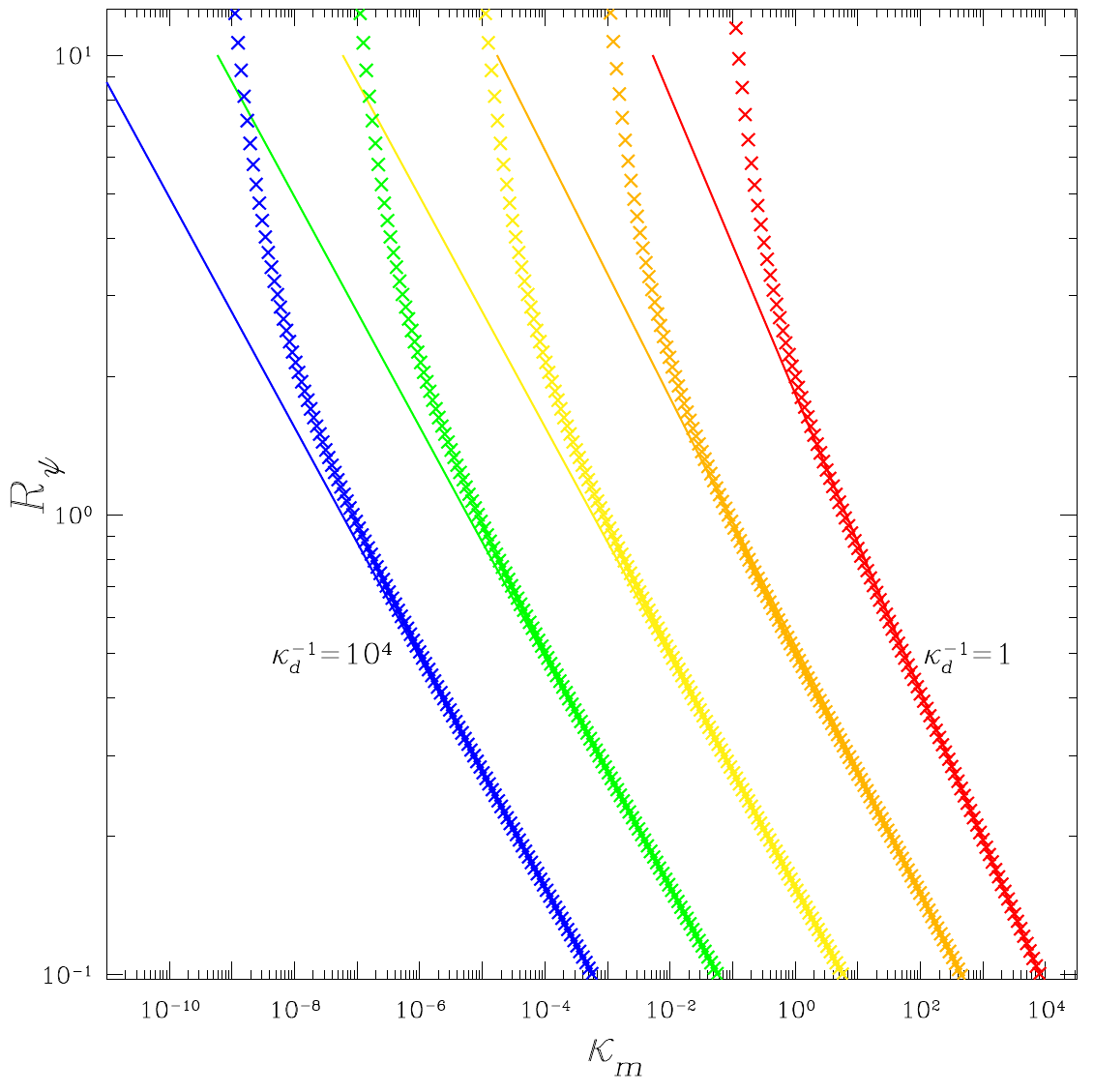}
\caption{
Plots of fermion bound state radius $R_\psi$ as a function of 
boson-fermion interaction strength, $\kappa_m$, for   
$\kappa_d^{-1}=1, 10, 10^2, 10^3, 10^4$ in 
red, orange, yellow, green, and blue x's, respectively.
The solid lines represent best-fit approximations to the small-$R$ solutions
that are in good agreement with the approximate closed-form 
solutions (see Table \ref{table:RpsiBestFit}).
 } 
\label{fig:Rd_vs_kdm1.pdf}
\end{center}
\end{figure}
%%%%%%%%%%%%%%%%%%%%%%%%%%%%%%%%%%%%%%%%%
%%%%%%%%%%%%%%%%%%%%%%%%%%%%%%%%%%%%%%%%%%%%%%%%%%%%%%%
\begin{table}[b]
%\begin{tabular}{S[table-format=5.3]  c S[table-format=5.4] }
\begin{tabular}{l l c }
\hline
\hline
$\kappa_d^{-1}$ & $b$ & $p$  \\
\hline
$1$		& 	$1.8$				& -0.32 \\
$10$		&	$5.2\times 10^{-1}$		& -0.27 \\
$10^2$	& 	$1.5\times 10^{-1}$		& -0.25 \\
$10^3$	& 	$4.9\times 10^{-2}$		& -0.25 \\
$10^4$	& 	$1.5\times 10^{-2}$		& -0.25 \\
\hline
\hspace{12mm} & \hspace{12mm} & \hspace{12mm} \\
\vspace{-7mm}
\end{tabular}
\caption{
Table of best-fit parameters for relationship $R_\psi \approx b \left(\kappa_m\right)^p$ for a range fermion effective 
masses, $\kappa_d^{-1}$.  
When comparing the numerically calculated  fermion radius $R_\psi$ to the width of the closed-form solutions, $\sigma$,
the numerically obtained solutions are in close agreement with the 
closed-form approximate solutions (\ref{eqn:TrapR2Relation}) 
that  predict a  $R_\psi \propto  \left(\kappa_m\right)^{-1/4}$ relationship, especially for larger effective fermion mass.
 % and are accurate to a few percent.
%
\label{table:RpsiBestFit}
}
\end{table}
%%%%%%%%%%%%%%%%%%%%%%%%%%%%%%%%%%%%%%%%%%%%%%%%%%%%%%%

\begin{figure}[]
% source location: ~/cpp/cv1dx/figs/Rd_vs_kdm1
\begin{center}
\includegraphics[width=80mm]{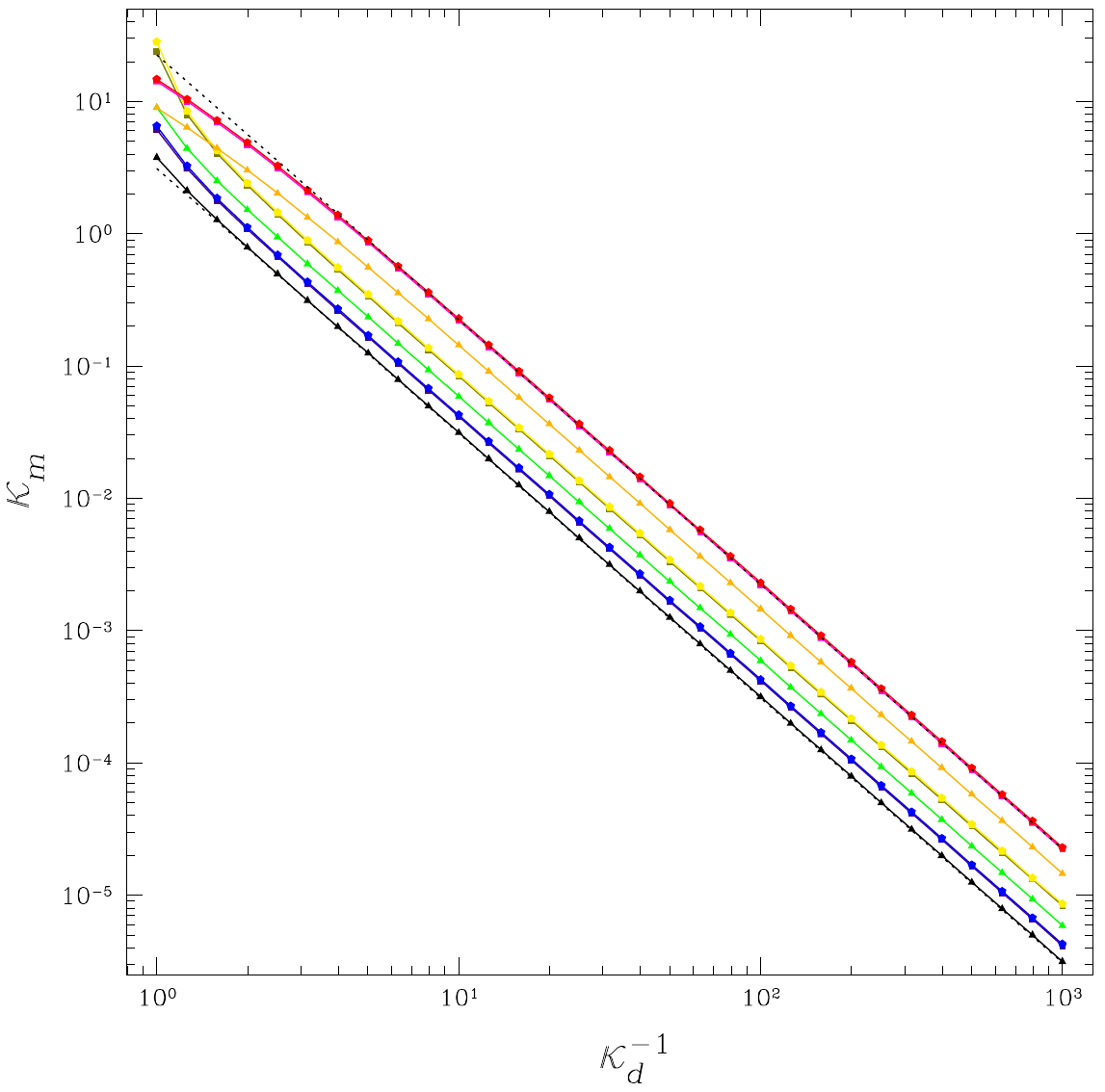}
\caption{
Plots of boson-fermion interaction strength, $\kappa_m$ required to 
keep a fermion bound state of mass $\kappa_d^{-1}$ confined to a 
particular radius, $R_\psi$. 
From top to bottom, 
plots for $R_\psi=0.75$ and $\kappa=100, 10, 1$ are in red, magenta, and orange, respectively;
plots for $R_\psi=1$ and $\kappa=100, 10, 1$ are in yellow, olive, and green, respectively; 
and
plots for $R_\psi=1.25$ and $\kappa=100, 10, 1$ are in blue, purple, and black, respectively. 
Plots are bounded above and below  by curves demonstrating $\kappa_m\propto (\kappa_d^{-1})^{-2}$ 
(dotted gray).
} 
\label{fig:km_vs_kdm1.pdf}
\end{center}
\end{figure}

The volume of parameter space where  bound states exist  is first explored by varying the boson-fermion interaction 
strength ($\kappa_m$). 
Figure \ref{fig:Rd_vs_kdm1.pdf} displays plots of the fermion bound state radius, $R_\psi$,  
as a function of $\kappa_m$ across  a range of effective fermion masses, $\kappa_d^{-1}$. 
The results show that increasing the interaction strength decreases the radius of a given bound state;
each curve (distinct $\kappa_d^{-1}$) is observed  to approach a  small $R_\psi$ limiting behavior  
of $R_\psi  \propto b \left(\kappa_m\right)^{-1/4}$ with best-fit parameters captured in Table \ref{table:RpsiBestFit}.
This is precisely the behavior predicted by the condition (\ref{eqn:TrapR2Relation}) on the 
closed form approximate solution (\ref{eqn:PsiGauHarmonicTrap}) that assumes the vortex acts as a 
 spatially harmonic trap.
However, as the interaction strength is decreased, the radius of the bound state increases, eventually extending 
beyond the size of the  trap.  As $\kappa_m$ is decreased even further the trap can no longer contain 
the fermion field, and bound states cease to exist. % for finite $\kappa_m$.
Although $\kappa_m$ can become arbitrarily small for arbitrarily large $\kappa_d^{-1}$, for any specific 
value of $\kappa_d^{-1}$ there is a non-zero finite $\kappa_m$ below which bound states do not exist. 
The roughly log-periodic spacing of the plots with different $\kappa_d^{-1}$ and the best fit values of $b$ 
captured in
Table \ref{table:RpsiBestFit} suggest that for a given $R_\psi$, the amount of interaction strength
needed to confine the fermion bound state scales as $\kappa_m \propto \kappa_d^2$.
To more definitively determine this relationship, \hbox{Figure \ref{fig:km_vs_kdm1.pdf} }
presents $\kappa_m$ as a function of $\kappa_d^{-1}$ for a range of GL parameters ($\kappa$) and selected fermion bound
state radii ($R_\psi$).
For each $\kappa$ and $R_\psi$ a curve is generated by fixing $\kappa_d^{-1}$  and  varying the interaction strength $\kappa_m$
until the radius of the resulting bound state is the desired $R_\psi$. 
The curves are shown to very closely demonstrate the expected  $\kappa_m \propto \kappa_d^2$ relationship.

%%%%%%%%%%%%%%%%%%%%%%%%%%%%%%%%%%%%%%%%%
%%%%%%%%%%%%%%%%%%%%%%%%%%%%%%%%%%%%%%%%%

%%%%%%%%%%%%%%%%%%%%%%%%%%%%%%%%%%%%%%%%%
\begin{figure}[]
% source location: ~/cpp/cv1dx/figs/Rd_vs_kdm1
\begin{center}
\includegraphics[width=80mm]{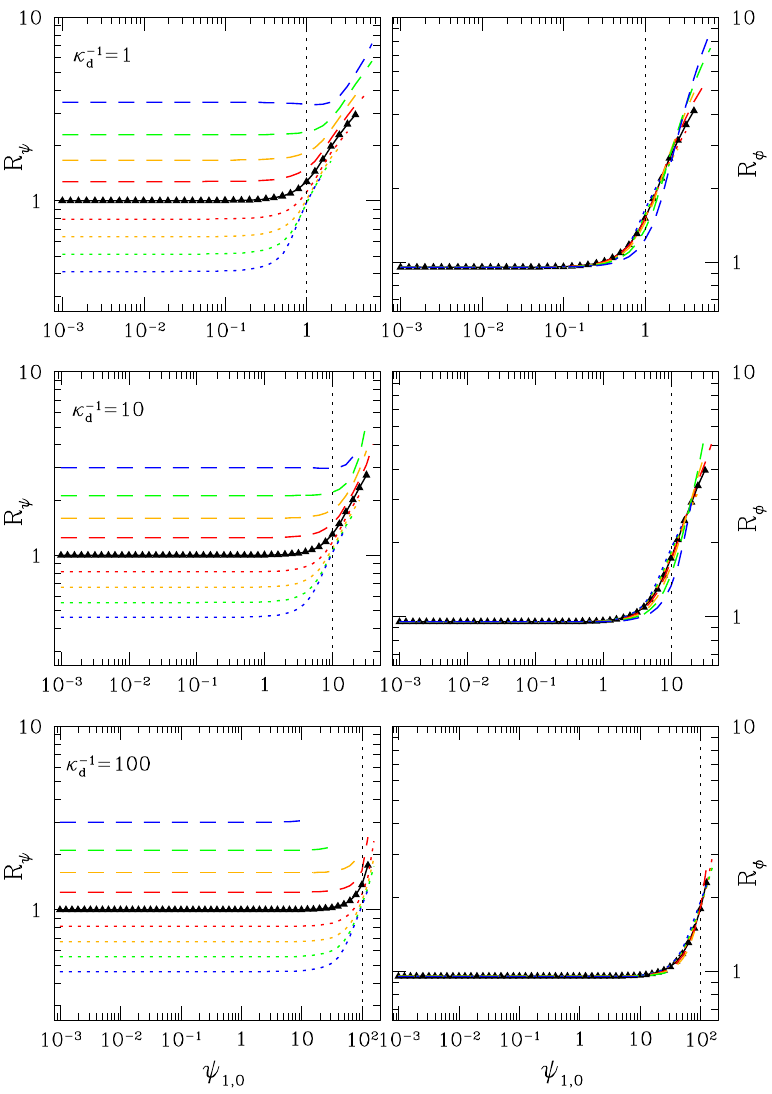}
\caption{ 
Plots of fermion bound state radius, $R_\psi$, and vortex core radius, $R_\phi$, 
as a function of fermion field strength for $\kappa_d^{-1}=1, 10, 100$ (top, middle, 
and bottom, respectively).  Each plot contains a curve (black) with the boson-fermion interaction 
strength $\kappa_m$ that confines the fermion bound state to $R_\psi = 1$ for field strength 
$\psi_{1,0} = 1 \times 10^{-3}$.  Additional solutions are provided by increasing $\kappa_m$ 
by a factor of 2, 4, 8, and 16 (dotted red, orange, green, and blue, respectively) and decreasing $\kappa_m$ 
by a factor of 2, 4, 8, and 16 (dashed red, orange, green, and blue, respectively).
A $\psi_{1,0} = \kappa_d^{-1}$ vertical dotted line is drawn on each graph and
all plots have GL parameter $\kappa=100$.
} 
\label{fig:RdRv_vs_Psi.pdf}
\end{center}
\end{figure}
%%%%%%%%%%%%%%%%%%%%%%%%%%%%%%%%%%%%%%%%%

Collectively, these results clearly demonstrate the existence of fermion bound states 
within the core of a vortex  across a wide range of the $(\kappa, \kappa_d^{-1}, \kappa_m)$ 
parameter space.  
It has been shown that when the bound state is sufficiently confined to the core of the vortex,  
the solutions closely resemble the closed-form   Gaussian solutions confined by a spatially harmonic trap.
It should be noted, however, that up to this point all solutions were generated with sufficiently low
fermion field strength as   not to significantly adjust the spontaneously broken vacuum of the condensate.
Guided again by the closed form results of Section \ref{subsec:ClosedForm} and the scalar potential (\ref{eqn:VphiSSB}), 
one  expects a shift in the scalar vacuum and
a loss of containment as $\kappa_m\bar{\psi}\psi \rightarrow 1$,
which occurs when the fermion field  approaches $\psi_{1,0}^2 \approx \kappa_d^{-2}$.

Figure \ref{fig:RdRv_vs_Psi.pdf} plots the fermion bound state radius ($R_\psi$) and the measured vortex radius ($R_\phi$) 
as a function of fermion field strength for a range of effective fermion masses ($\kappa_d^{-1}$) 
and interaction strengths ($\kappa_m$).
Baseline values of $\kappa_m$ are determined for each that result in $R_\psi = 1$ in the limit of weak fermion field;
in this limit the measured vortex radius is close to unity and approaches the traditional (uncharged) vortex solution.
As  $\psi_{1,0}$ approaches $\kappa_d^{-1}$, the force between 
the fermion bound state and the confining scalar vortex becomes significant and the wall of the scalar vortex begins to expand
 so that the outward self-repulsion of the fermion field can balance with the inward force of the vortex wall. 
The vortex is able to accommodate some pressure, but as the fermion bound state expands beyond the vortex, 
sufficient confinement force is no longer present, and solutions to the stationary equations can no longer be obtained.
%
 
%%%%%%%%%%%%%%%%%%%%%%%%%%%%%%%%%%%%%%%%%
\begin{figure}[]
% source location: ~/cpp/cv1dx/figs/Rd_vs_kdm1
\begin{center}
\includegraphics[width=80mm]{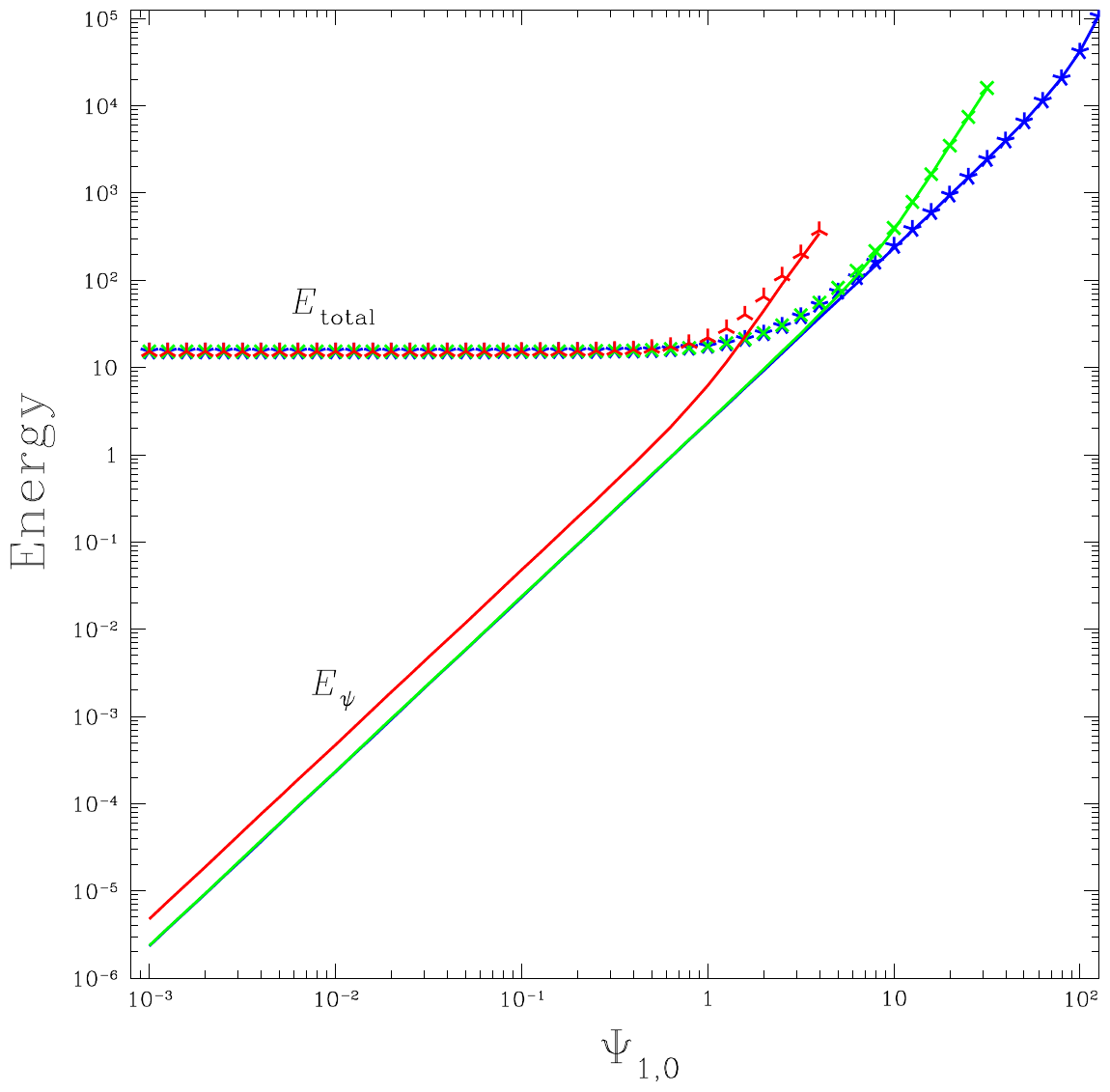}
\caption{
Plots of fermion energy (solid lines) and total energy (points) as a function of fermion field strength $\psi_{1,0}$
for $\kappa = 100$ and $\kappa_d^{-1}=1,10, 100$ in red, green, and blue, respectively. 
The fermion energy is observed to exceed the uncharged vortex energy when $\psi_{1,0} \gtrsim 1$,
and and for $\kappa_d^{-1}=100$, total energies exceeding $10^4$ times the uncharged energy were observed.
} 
\label{fig:Masses.pdf}
\end{center}
\end{figure}
%%%%%%%%%%%%%%%%%%%%%%%%%%%%%%%%%%%%%%%%%%%%%%%%%%%%%%%

While not explored parametrically in detail here, it is worth briefly discussing how the addition of a fermion bound state 
impacts the overall mass/energy of the charged vortex  solution.  
In Figure \ref{fig:Masses.pdf} one can see that for $\psi_{1,0}$ significantly less than unity the  
energy from the fermion field does not significantly contribute to the overall energy.   However, when the fermion field strength at the origin 
grows larger than unity, the contribution by the fermion bound state to the overall energy can become quite large.
 For solutions with $\kappa_d^{-1}=100$, the ratio of the fermion energy to an uncharged ($\psi=0$) vortex energy 
 was observed to  exceed $10^4$.

%%%%%%%%%%%%%%%%%%%%%%%%%%
\subsection{Time evolution and stability}
%%%%%%%%%%%%%%%%%%%%%%%%%%

%
%
Previous observations about the existence of stationary solutions were based on the ability to find a solution to the 
time-independent stationary equations of  motion; however, these solutions may not be long-term stable.
This section %subsection 
presents the time evolution of stationary charged vortices of Section \ref{subsec:Stationary} with the intent 
of better understanding their stability properties.
The time evolution equations for the scalar field are given 
by
\begin{eqnarray}
%%%%%%%%%%%%%%%%%%%%%%%%%%%%%%%%%%%%%
\partial_t \Pi_1 
&=& 
\partial_R \chi_{1} 
+ \frac{2}{\kappa}\left( -A_t\Pi_2 + A_R \left( \chi_2-\frac{\phi_2}{R}\right)  + \frac{1}{R}\tilde{A}_\theta \phi_1\right)
\nonumber \\ &&
- \kappa^{-2} \phi_1 \left( -A_t^2 + A_R^2 +\tilde{A}_\theta^2\right)
\nonumber \\ &&
+\phi_1\left(1-\kappa_m \left( a_1^2 + a_2^2 - b_1^2 - b_2^2\right)  \right) 
\nonumber \\  &&
- \phi_1\left( \phi_1^2 + \phi_2^2\right), 
\\
%\end{eqnarray}
%\begin{eqnarray}
%%%%%%%%%%%%%%%%%%%%%%%%%%%%%%%%%%%%%
\partial_t \Pi_2 
&=& 
\partial_R\chi_2
- \frac{2}{\kappa}\left( -A_t\Pi_1 + A_R\left( \chi_1-\frac{\phi_1}{R}\right) - \frac{1}{R}\tilde{A}_\theta\phi_2 \right)
\nonumber \\ &&
- \kappa^{-2} \phi_2 \left( -A_t^2 + A_R^2 + \tilde{A}_\theta^2   \right)
\nonumber \\ &&
+\phi_2\left(1-\kappa_m\left( a_1^2 + a_2^2 - b_1^2 - b_2^2\right) \right) 
\nonumber \\  &&
- \phi_2\left( \phi_1^2 + \phi_2^2\right), 
\\
%\end{eqnarray}
%\begin{eqnarray}
%%%%%%%%%%%%%%%%%%%%%%%%%%%%%%%%%%%%%
\partial_t\chi_1  &=& 2\frac{\partial}{\partial R^2} \left( R \Pi_1 \right), \\
%%%%%%%%%%%%%%%%%%%%%%%%%%%%%%%%%%%%%
\partial_t\chi_2  &=& 2\frac{\partial}{\partial R^2} \left( R \Pi_2 \right), \\
%%%%%%%%%%%%%%%%%%%%%%%%%%%%%%%%%%%%%
\partial_t\phi_1 &=& \Pi_1, \text{ and }
\\
%%%%%%%%%%%%%%%%%%%%%%%%%%%%%%%%%%%%%
\partial_t\phi_2 &=& \Pi_2. 
\end{eqnarray}
The evolution equations for the electromagnetic fields are 
\begin{eqnarray}
%%%%%%%%%%%%%%%%%%%%%%%%%%%%%%%%%%%%%
 \partial_t E^R
&=&
 \frac{1}{\kappa}  \left(  \phi_2 \chi_1-   \phi_1\chi_2  \right)
+ \kappa^{-2} A_R (\phi_1^2 + \phi_2^2)
\nonumber \\ &&
- \frac{\kappa_{d}}{\kappa}  \left( a_1b_1 + a_2b_2\right),
\\
%%%%%%%%%%%%%%%%%%%%%%%%%%%%%%%%%%%%%
\partial_t  \tilde{E}^\theta
&=&
- \partial_R  B^z 
+ \left(  \frac{\tilde{A}_\theta}{\kappa^2} - \frac{1}{\kappa R}\right)\left( \phi_1^2 + \phi_2^2 \right)
\nonumber \\ &&
-\frac{\kappa_{d}}{\kappa}  \left( a_1b_2 - a_2b_1  \right), 
\\
%%%%%%%%%%%%%%%%%%%%%%%%%%%%%%%%%%%%%
\partial_{t} B_z &=&  -\frac{1}{R}\partial_{R}  (R\tilde{E}_\theta),  \\
%%%%%%%%%%%%%%%%%%%%%%%%%%%%%%%%%%%%%
\partial_t A_t &=&  \frac{1}{R}\partial_R\left( R A_R \right),   \\
%%%%%%%%%%%%%%%%%%%%%%%%%%%%%%%%%%%%%
\partial_t A_R  &=&  \partial_R A_t -E_R, \text{ and} 
\\
%%%%%%%%%%%%%%%%%%%%%%%%%%%%%%%%%%%%%
\partial_t \tilde{A}_\phi   &=& -\tilde{E}_\theta.
\end{eqnarray}
And finally for the fermion field,
\begin{eqnarray}
%%%%%%%%%%%%%%%%%%%%%%%%%%%%%%%%%%%%%
\partial_t a_1  &=&
-  \frac{1}{R} \partial_R \left(R b_1 \right)
+\frac{1}{2\kappa}
\Big[ 
- A_t a_2  - A_R b_2   + \tilde{A}_\theta b_1  %+ A_z\Psi_3 
\Big] 
\nonumber \\ &&
+ \frac{1}{\kappa_d}  \left(1+\frac{1}{2}\kappa_m\hat{\phi}^2 \right) a_2,
\\
%%%%%%%%%%%%%%%%%%%%%%%%%%%%%%%%%%%%%
\partial_t a_2  &=&
-  \frac{1}{R} \partial_R \left(R b_2 \right)
+\frac{1}{2\kappa} 
\Big[ 
A_t a_1  +  A_R b_1   + \tilde{A}_\theta b_2 %+ A_z\Psi_3 
\Big] 
\nonumber \\ &&
- \frac{1}{\kappa_d}  \left(1+\frac{1}{2}\kappa_m\hat{\phi}^2 \right)a_1,  
\\
%%%%%%%%%%%%%%%%%%%%%%%%%%%%%%%%%%%%%
\partial_t b_1 &=&
-\partial_R  a_1
+\frac{1}{2\kappa}
\Big[ 
-A_t b_2  - A_R a_2   - \tilde{A}_\theta a_1 %- A_z\Psi_2
\Big] 
\nonumber \\ &&
-   \frac{1}{\kappa_d}  \left(1+\frac{1}{2}\kappa_m\hat{\phi}^2 \right)b_2,  \text{ and}
\\
%%%%%%%%%%%%%%%%%%%%%%%%%%%%%%%%%%%%%
\partial_t b_2 &=&
-\partial_R  a_2
+\frac{1}{2\kappa}
\Big[ 
A_t b_1  + A_R a_1   - \tilde{A}_\theta a_2 %- A_z\Psi_2
\Big] 
\nonumber \\ &&
+   \frac{1}{\kappa_d}  \left(1+\frac{1}{2}\kappa_m\hat{\phi}^2 \right)b_1.
%%%%%%%%%%%%%%%%%%%%%%%%%%%%%%%%%%%%%
%%%%%%%%%%%%%%%%%%%%%%%%%%%%%%%%%%%%%
\end{eqnarray}
\begin{figure}[]
% source location: ~/cpp/cv1dx/figs/Rd_vs_kdm1
\begin{center}
\includegraphics[width=85mm]{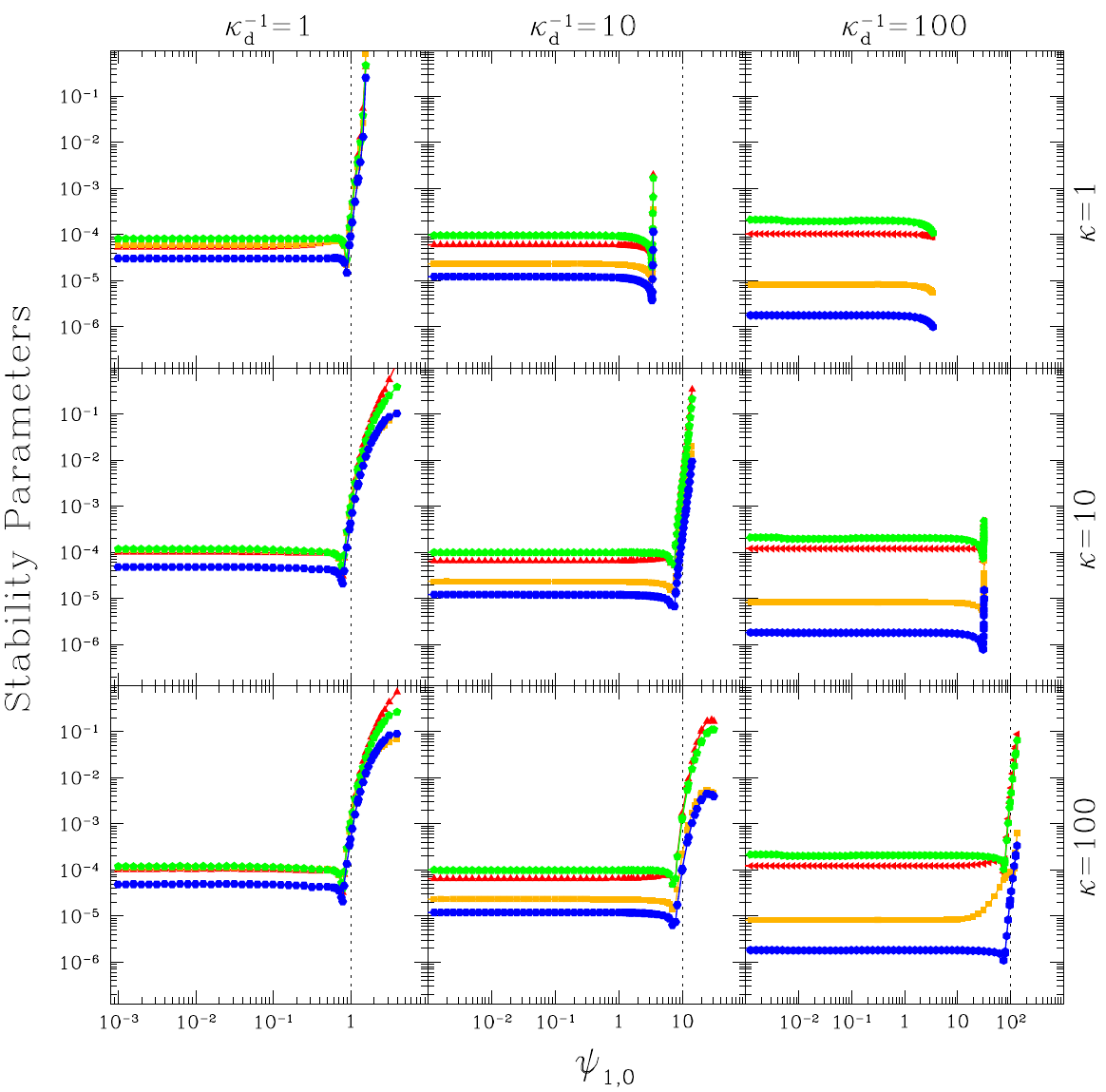}
\caption{ 
Plots of stability parameters as a function of fermion field strength $\psi_{1,0}$ 
for a range of $\kappa_d^{-1}$ and $\kappa$.
The L2-norms of the difference in max/min envelopes for $|\psi_1|$ and $|\psi_4|$ normalized by $\psi_{1,0}$ 
are plotted in  green and blue, respectively.
The standard deviation in the fermion radius, $R_\psi$, is plotted in red,
and the standard deviation in the frequency of the fermion field at  $R=0$ is plotted in orange.
Evolutions with $\kappa_d^{-1}=100$ were evolved to $t=200$, and all other evolutions to $t=40$,
and a $\psi_{1,0} = \kappa_d^{-1}$ vertical dotted line is drawn on each graph.
} 
\label{fig:Stability.pdf}
\end{center}
\end{figure}
Figure \ref{fig:Stability.pdf} displays four stability parameters obtained by time evolving 
solutions to the stationary equations of motion for a wide range of GL parameters ($\kappa$) and 
effective fermion mass ($\kappa_d^{-1}$);
each parameter measures the variation in a quantity that should remain constant over time for 
a stable bound state.
The first stability parameter considered is the standard deviation of radius of the bound state ($R_\psi$)
over  the duration of the simulation.  
Also considered is the standard deviation of the instantaneous angular frequency of the $\psi_1$ component
of the fermion field
\begin{eqnarray}
\omega &=& \left( a_2\partial_ta_1 - a_1\partial_ta_2 \right) / \left(a_1^2 + a_2^2\right)
\end{eqnarray}
at the core of the vortex.
The last two stability parameters are obtained by calculating the  difference between the maximum and minimum of 
$|\psi_1(t,R)|$ and $|\psi_4(t,R)|$ over time at each point and then taking the L2-norm over a spatial 
domain containing the vortex and bound state.  
The results show that bound state solutions are stable when  $\bar{\psi}\psi$ does not significantly 
change the vacuum of the scalar field.
For small $\bar{\psi}\psi$  the trap remains harmonic, but 
when $\bar{\psi}\psi\rightarrow \kappa_m^{-1}$, the harmonic trap deforms and flattens out.
For $\kappa \gtrsim \kappa_d^{-1}$, 
the fermion bound states are observed to be stable up to $\psi_{1,0}\approx\kappa_d^{-1}$ as suggested by the closed-form analysis.  
For $\kappa \ll \kappa_d^{-1}$,  time-independent stationary bound states were not even obtainable up to the $\psi_{1,0}\approx\kappa_d^{-1}$ threshold, 
but the stationary states that do exist  are observed to be stable up to that threshold, with a small region of instability on the boundary.
When $\bar{\psi}\psi$ is large, two distinct effects  lead to instability.  
First, the boson-fermion interaction becomes strong enough to change the shape
of the trap (\ref{eqn:VphiSSB}), flattening  the trap and lowering the confinement force on the bound state.
Second, for $\kappa \ll \kappa_d^{-1}$, the gauge potential influences the effective potential (\ref{eqn:EffectivePotential})
by reducing the potential barrier that would confine the bound state, thereby lowering the fermion energy
that can be supported by the trap.

%%%%%%%%%%%%%%%%%%%%%%%%%%%%%%%%%%%%%%%%%%%%%%%%%%%%%%%%%%%%%
%%%%%%%%%%%%%%%%%%%%%%%%%%%%%%%%%%%%%%%%%%%%%%%%%%%%%%%%%%%%%
\section{Head-on Scattering \label{sec:Scattering}}
%%%%%%%%%%%%%%%%%%%%%%%%%%%%%%%%%%%%%%%%%%%%%%%%%%%%%%%%%%%%%
%%%%%%%%%%%%%%%%%%%%%%%%%%%%%%%%%%%%%%%%%%%%%%%%%%%%%%%%%%%%%

Previous sections have established the conditions for the existence and stability of fermion bound states within
a vortex core when a repulsive boson-fermion interaction is present.
This section offers a brief look at the scattering of two such charged vortices.
For each simulation, initial data are  created using the time-independent stationary equations of motion to generate 
at-rest vortex solutions.  The at-rest solutions are individually boosted head-on at one another (zero impact parameter)
and superimposed to create the initial data.
The time-dependent (2+1) equations of motion are then evolved to simulate the scattering event.

The boost equations are described in Appendix \ref{app:NumericalMethods}, while 
the time-evolution equations  for the complex scalar field are given by
\begin{eqnarray}
\partial_t\Pi_1 
&=&
\partial_x\Phi_{1x} + \partial_y\Phi_{1y}  
-\kappa^{-2} \phi_1 \left( -A_t^2 + A_x^2 + A_y^2 \right)  
\nonumber \\ &&
+2\kappa^{-1} \left(  -A_t \Pi_2 + A_x\Phi_{2x}+ A_y\Phi_{2y}  \right)
\nonumber \\ &&
+\phi_1\left(1-\kappa_m \left(a_1^2 + a_2^2 - b_1^2 - b_2^2\right) \right)   
\nonumber \\ &&
- \phi_1\left(\phi_1^2 + \phi_2^2\right), 
\label{eqn:First2p1Evol}
 \\
%%%%%%%%%%%%%%%%%%
\partial_t\Pi_2
&=&
\partial_x\Phi_{2x} + \partial_y\Phi_{2y}  
-\kappa^{-2} \phi_2 \left( -A_t^2 + A_x^2 + A_y^2 \right)  
\nonumber \\ &&
-2\kappa^{-1} \left(  -A_t \Pi_1 + A_x\Phi_{1x}+ A_y\Phi_{1y}  \right)
\nonumber \\ &&
+\phi_2\left(1-\kappa_m    \left(a_1^2 + a_2^2 - b_1^2 - b_2^2\right)\right) 
\nonumber \\ &&
- \phi_2\left(\phi_1^2 + \phi_2^2\right),
 \\
%%%%%%%%%%%%%%%%%%
\partial_t\Phi_{1x} &=& \partial_x\Pi_1, \\
\partial_t\Phi_{1y} &=& \partial_y\Pi_1, \\
\partial_t\Phi_{2x} &=& \partial_x\Pi_2, \\
\partial_t\Phi_{2y} &=& \partial_y\Pi_2, \\
\partial_t\phi_1 &=& \Pi_1, \text{ and}\\
\partial_t\phi_2 &=& \Pi_2.
\end{eqnarray}
%%%%%%%%%%%%%%%%%%
%
The equations for the electromagnetic fields and vector potential are given by
\begin{eqnarray}
%%%%%%%%%%%%%%%%%%
\partial_t E_x   
&=&
 \partial_yB_z
 +\frac{1}{\kappa} \left(  \phi_2 \Phi_{1x}  -   \phi_1 \Phi_{2x}  \right)
+ \kappa^{-2}A_x \left(\phi_1^2 + \phi_2^2\right)
\nonumber \\ &&
- \kappa_{d} \kappa^{-1} \left(a_1b_1 + a_2b_2 \right),
\\
\partial_tE_y  
&=&
-\partial_xB_z
 +\frac{1}{\kappa}    \left(  \phi_2  \Phi_{1y}  -   \phi_1  \Phi_{2y}  \right)
+ \kappa^{-2}A^y \left(\phi_1^2 + \phi_2^2\right)
\nonumber \\ &&
- \kappa_{d} \kappa^{-1} \left(a_1b_2 - a_2b_1 \right), 
\\
\partial_tE_z 
&=&
 \partial_xB_y - \partial_yB_x,
%- \frac{1}{4}\kappa_{d} \kappa^{-1} \left( \bar{\psi}\gamma^z \psi  \right)
\\%\end{eqnarray}
%\begin{eqnarray}
%%%%%%%%%%%%%%%%%%
\partial_t  B_x
%&=& 
%-\left(\partial_y E_z
%-\partial_z  E_y \right)
&=& 
-\partial_y E_z,
\\
\partial_t  B_y
%&=& 
%-\left( 
%\partial_z  E_x
%-
%\partial_x  E_z
%\right)
&=& 
\partial_x  E_z,
\\
\partial_{t}  B_z
&=& 
-\left(
\partial_{x}  E_y 
-
\partial_{y}  E_x 
\right),
%%%%%%%%%%%%%%
\\%\end{eqnarray}
%\begin{eqnarray}
%%%%%%%%%%%%%%
\partial_tA_x &=& \partial_xA_t -E_x, \\
%%%
\partial_tA_y &=& \partial_yA_t -E_y,  \text{ and}\\
%%%
\partial_t A_t &=&\partial_x A_x + \partial_y A_y.
\end{eqnarray}
%%%%%%%%%%%%%%%%%%
%
And finally, the evolution equations for the fermion field components are given by
\begin{eqnarray}
%
%%%%%%%%%%%%%%%%%%%%%%%%%%%%%%
 \partial_ta_1  &=&
 -  \partial_xb_1   - \partial_yb_2
+\frac{1}{2}\kappa^{-1}\left(
-A_t a_2 
-A_xb_2 + A_yb_1
\right)
\nonumber \\ &&
%+\kappa_d^{-1}\mathcal{M}(\phi) a_2
+\kappa_d^{-1}  \big(1+\frac{1}{2}\kappa_m \left(\phi_1^2 + \phi_2^2\right) \big) a_2,
\\
%%%%%%
 \partial_t a_2 &=&
 -\partial_xb_2 + \partial_yb_1
+\frac{1}{2}\kappa^{-1}\left(
A_ta_1  + A_xb_1 + A_yb_2
\right)
\nonumber \\ &&
%-\kappa_d^{-1}\mathcal{M}(\phi) a_1 
-\kappa_d^{-1} \big(1+\frac{1}{2}\kappa_m \left(\phi_1^2 + \phi_2^2\right) \big) a_1, 
\\
%%%%%%%%%%%%%%%%%%%%%%%%%%%%%%
 \partial_tb_1    &=&
-\partial_xa_1 + \partial_ya_2
+\frac{1}{2}\kappa^{-1}\left(
-A_t b_2  - A_xa_2  - A_ya_1 
\right)
\nonumber \\ &&
%-\kappa_d^{-1}\mathcal{M}(\phi) b_2
-\kappa_d^{-1} \big(1+\frac{1}{2}\kappa_m \left(\phi_1^2 + \phi_2^2\right) \big) b_2, \text{ and}
\\
%%%%%%%%%%%%%%%%%%%%%%%%%%%%%%
 \partial_t b_2  &=&
 -\partial_xa_2 - \partial_ya_1
+\frac{1}{2}\kappa^{-1}\left(
A_t b_1  + A_xa_1 -A_ya_2
\right)
\nonumber \\ &&
%+\kappa_d^{-1}\mathcal{M}(\phi) b_1
+\kappa_d^{-1} \big(1+\frac{1}{2}\kappa_m \left(\phi_1^2 + \phi_2^2\right) \big) b_1.
\label{eqn:Last2p1Evol}
\end{eqnarray}
%%%%%%%%%%%%%%%%%%
%
%
The fact that vortices scatter at right angles when colliding head-on and above a critical velocity is well known
and discussed throughout the literature 
\cite{SHELLARD_1988262,
Myers_PhysRevD.45.1355,
RUBACK_1988669,
Manton_Topological_solitons_2004tk,
Honda_PhysRevD.102.056011}.
This work focuses on new phenomenology that arises with the addition of a fermion bound state.
As  an in-depth scattering analysis will be saved for future work, this initial look at scattering is limited to 
effective fermion mass $\kappa_d^{-1}=1$ to keep the frequency of oscillation of the fermion field on the order of unity, 
and $\kappa=2$ to allow a transition (from back- to right-angle scattering) boost velocity on the order 
of $v_b\approx 0.5$; such a velocity allows the dynamics to unfold quickly while also not significantly Lorentz contracting 
the solutions and requiring very fine numerical grids \cite{Honda_PhysRevD.102.056011}.
%
%
%
%
%%%%%%%%%%%%%%%%%%%%%%%%%%%%%%%%%%%%%%%%%%%%%%%%%%%%%%%%%%%%%%
\begin{figure}[]
\begin{center}
\includegraphics[width=80mm]{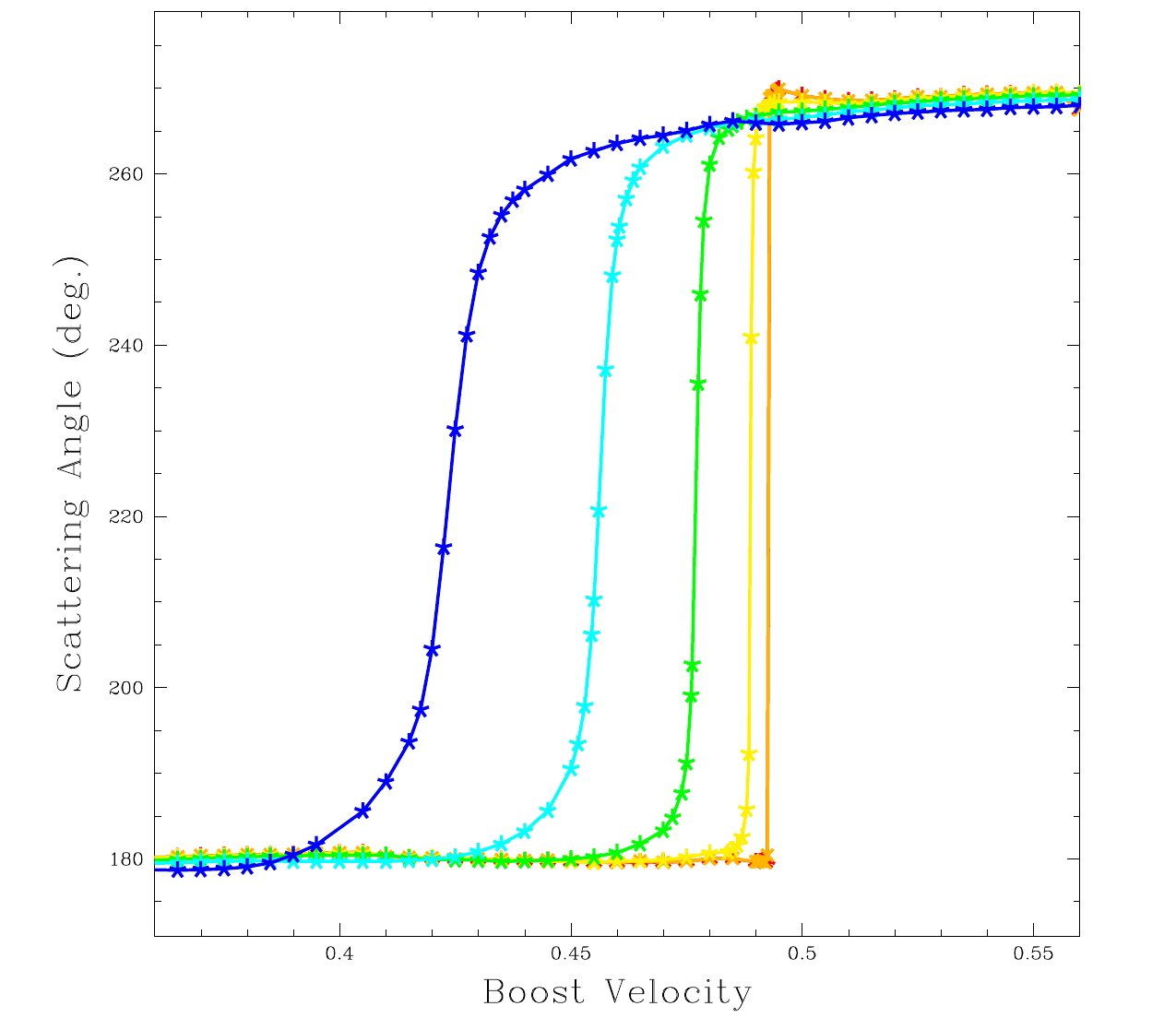}
\caption{ 
Plots of scattering angle as a function of boost velocity for head-on collisions of identical vortices.
Curves are presented for fermion field strengths of $\psi_{1,0} = 10^{-3}, 10^{-2},  10^{-1}, 2\times 10^{-1}, 3\times 10^{-1}, 4\times 10^{-1}$  
in red, orange, yellow, green, cyan, and blue, respectively.
Vortices are boosted at each other with zero initial impact parameter. 
All solutions are for $\kappa_d^{-1}=1$, $\kappa=2$, and use a $\kappa_m$ that gives $R_\psi=1$ for $\psi_{1,0}=10^{-3}$.
} 
\label{fig:RAS_vs_Psi10.pdf}
\end{center}
\end{figure}
%%%%%%%%%%%%%%%%%%%%%%%%%%%%%%%%%%%%%%%%%%%%%%%%%%%%%%%%%%%%%%
%%%%%%%%%%%%%%%%%%%%%%%%%%%%%%%%%%%%%%%%%%%%%%%%%%%%%%%%%%%%%%
\begin{figure}%~/cpp/bfh1d2d/PROD_WeakField2/*.jpg
\def \TILESCALE {1.0}
\includegraphics[width=\TILESCALE\linewidth]{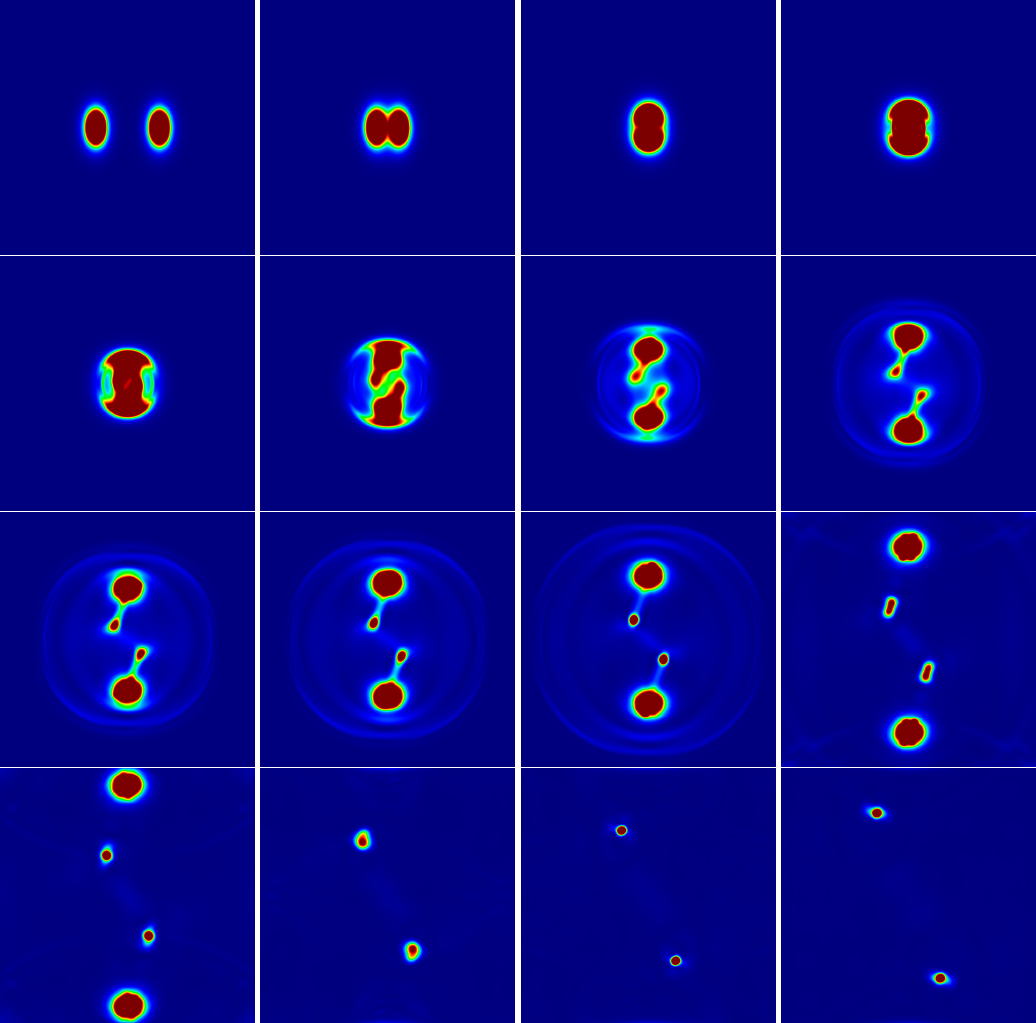}
\caption{
Time slices of the energy density for the head-on collision of two charged vortices with $v_x(0) = \pm 0.75$ 
and $\psi_{1,0}=0.4$.
The plots represent times for $n=$ 0, 175, 270, 315, 365, 420, 500, 600, 750, 900, 1000, 1400, 1650, 2100, 2400,
and  2944,   
where $t=n \Delta t$ and $\Delta t = 0.025$,
displayed from top left to bottom right, respectively.
The displayed simulations were conducted on an $801\times 801$ grid with domain spanning 
$\{x:-20<x<20\}$ and 
$\{y:-20<y<20\}$.
The sequence  shows approximate right-angle scattering of the charged vortices and  the generation of 
two stable bubbles.
\label{fig:headon_bp}}
\end{figure}
%%%%%%%%%%%%%%%%%%%%%%%%%%%%%%%%%%%%%%%%%%%%%%%%%%%%%%%%%%%%%%
%%%%%%%%%%%%%%%%%%%%%%%%%%%%%%%%%%%%%%%%%%%%%%%%%%%%%%%%%%%%%%
\begin{figure}% ~/cpp/bfh1d2d/PROD_WeakField/*.jpg
\def \TILESCALE {1.0}
\includegraphics[width=\TILESCALE\linewidth]{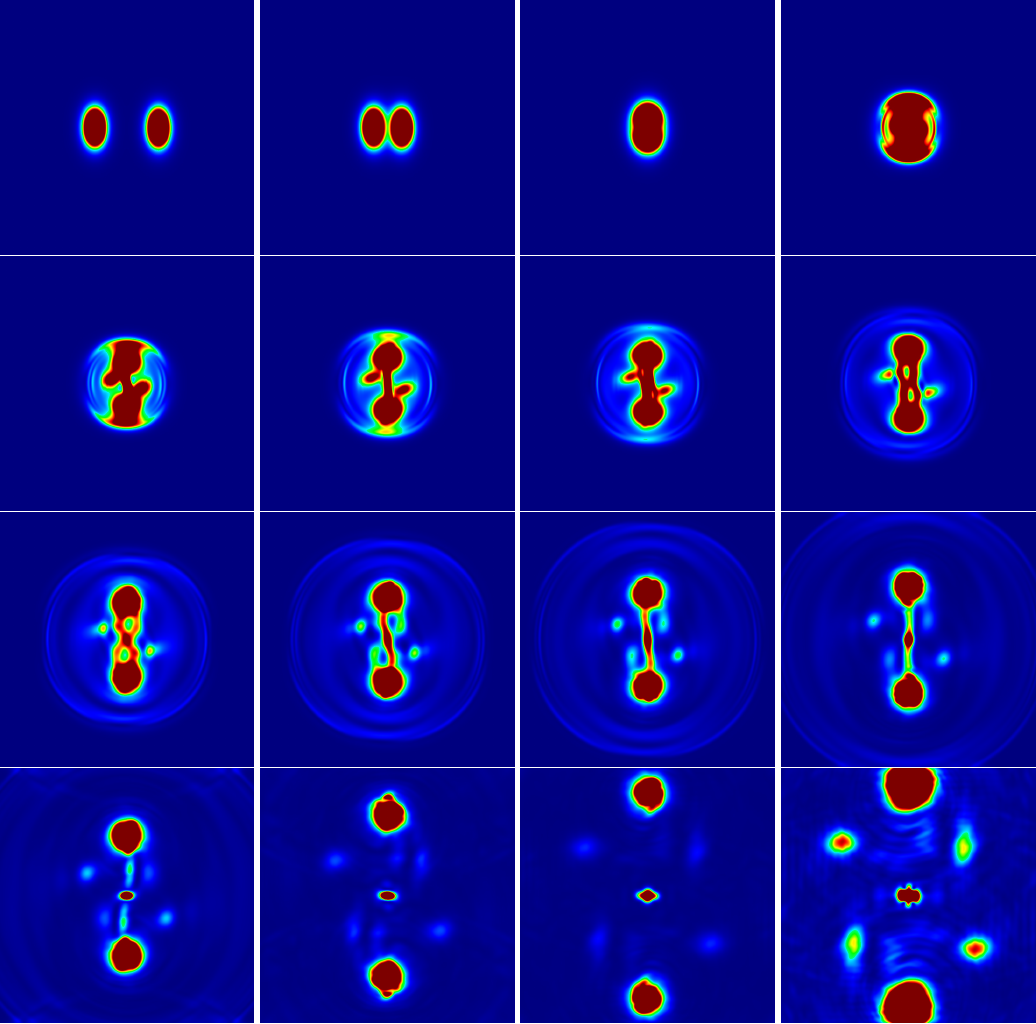}
\caption{
Time slices of the energy density for the head-on collision of two charged vortices with $v_x(0) = \pm 0.75$ 
and $\psi_{1,0}=0.5$.
The plots represent times for $n=$ 0, 150, 300, 425,  500, 557, 600, 700,   800, 900, 1000, 1150,  1300, 1800, and 2300, 
where $t=n \Delta t$ and $\Delta t = 0.025$,
displayed from top left to bottom right, respectively.
The displayed simulations were conducted on an $801\times 801$ grid with domain spanning 
$\{x:-20<x<20\}$ and 
$\{y:-20<y<20\}$.
The sequence  shows approximate right-angle scattering of the charged vortices, the generation of an at-rest bubble at the 
system center of mass,  and  the generation of four bubbles being ejected outward from the collision.
%two stable bubbles.
%
The colormap for the plot at $n=2500$ was rescaled to more clearly see the outgoing bubbles.
\label{fig:headon_bp2}}
\end{figure}
%%%%%%%%%%%%%%%%%%%%%%%%%%%%%%%%%%%%%%%%%%%%%%%%%%%%%%%%%%%%%%
%%%%%%%%%%%%%%%%%%%%%%%%%%%%%%%%%%%%%%%%%%%%%%%%%%%%%%%%%%%%%%
\begin{figure}%[h]PhiRho
\def \TILESCALE {1.0}
\includegraphics[width=\TILESCALE\linewidth]{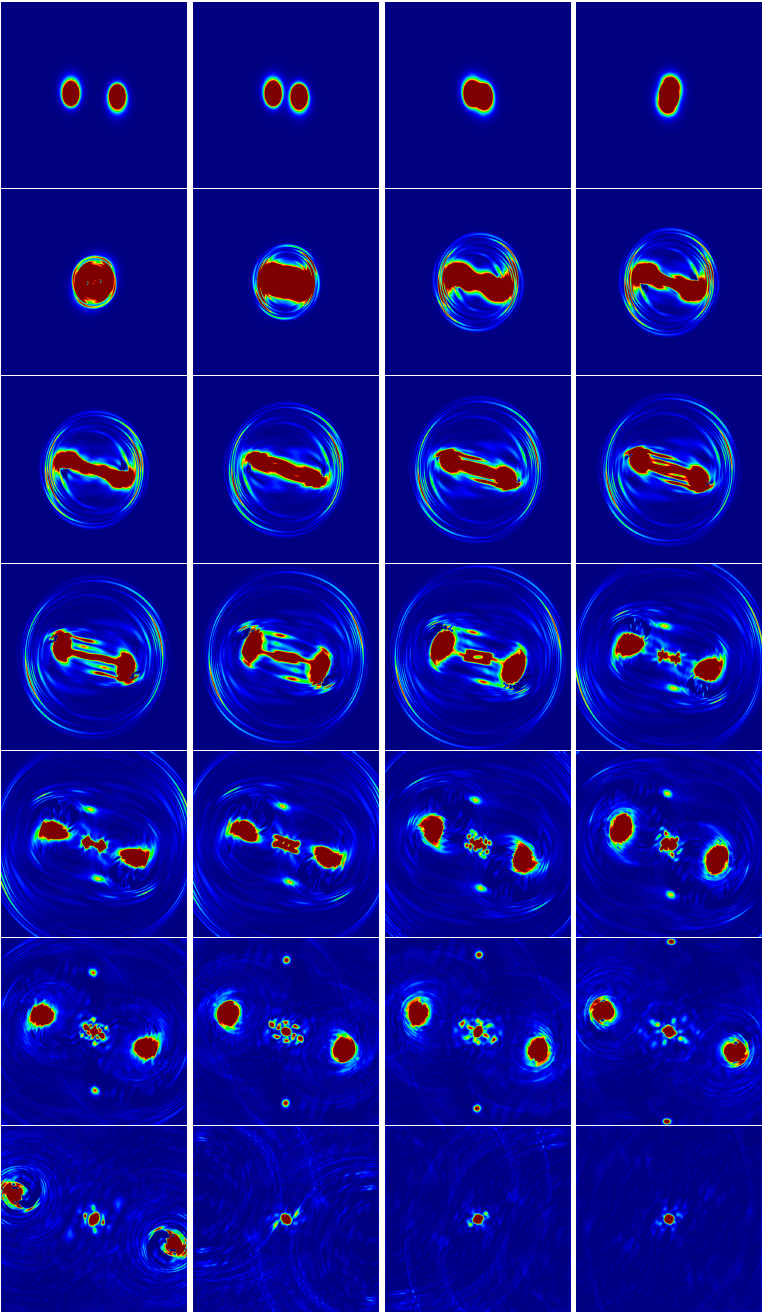}
\caption{
Time slices of the energy density for the head-on collision of two charged vortices with $v_x(0) = \pm 0.75$
and $\psi_{1,0}=1.5$.
The plots represent times for $n=$ 0, 120, 200, 290, 400, 500, 600, 640, 680, 750, 800, 825,
875, 950, 1000, 1100, 1150, 1200, 1300, 1400, 1526, 1720, 1800, 2000, 2500, 3200, 3600,  and 4000, 
where $t=n \Delta t$ and $\Delta t = 0.025$,
displayed from top left to bottom right, respectively.
The displayed simulations were conducted on an $801\times 801$ grid with domain spanning 
$\{x:-20<x<20\}$ and 
$\{y:-20<y<20\}$.
%
%The solution shows rich nonlinear inelastic scattering behaviors, including stable and transient 
%bubble production.
%
%
The sequence  shows rich  inelastic interactions that eject the vortices at off-angles (not back or
right-angle scattering), the creation and ejection of two bubbles, and the creation of a stationary 
bubble at the system center of mass.
\label{fig:headon_bp3}}
\end{figure}
%
%
%%%%%%%%%%%%%%%%%%%%%%%%%%%%%%%%%%%%%%%%%%%%%%%%%%%%%%%%%%%%%%
\begin{figure}%[h]PhiRho
\def \TILESCALE {1.0}
\includegraphics[width=\TILESCALE\linewidth]{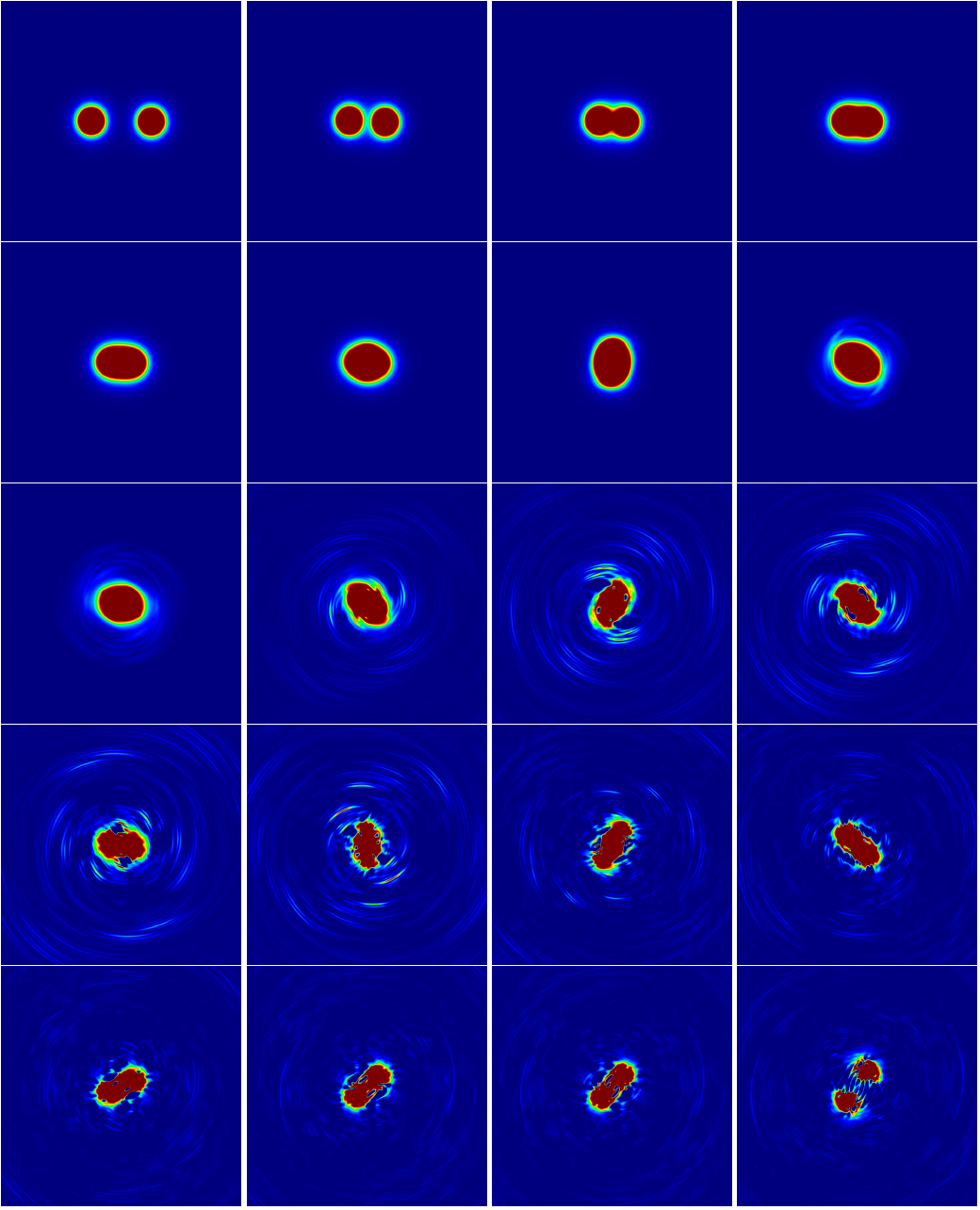}
\caption{
Time slices of the energy density for the head-on collision of two charged vortices with $v_x(0) = \pm 0.25$
and $\psi_{1,0}=1.0$.
The plots represent times for $n=$ 0, 350, 550,   682, 725, 797,    972, 1132, 1222, 
5467, 5930, 6173,    6337, 6754, 7981, 8262,    8579, 8665, 8693, and 8810 
where $t=n \Delta t$ and $\Delta t = 0.025$,
displayed from top left to bottom right, respectively.
The displayed simulations were conducted on an $801\times 801$ grid with domain spanning 
$\{x:-20<x<20\}$ and 
$\{y:-20<y<20\}$.
%
%The solution shows rich nonlinear inelastic scattering behaviors, including stable and transient 
%bubble production.
%
%
The sequence  shows coalescence around $n\approx 600$, stable interaction through $n\approx 4000$, 
unstable %turbulent 
dynamics through $n\approx 8600$, followed by eventual dispersal.
\label{fig:headon_bp4}}
\end{figure}
\begin{figure}[]
\begin{center}
\includegraphics[width=70mm]{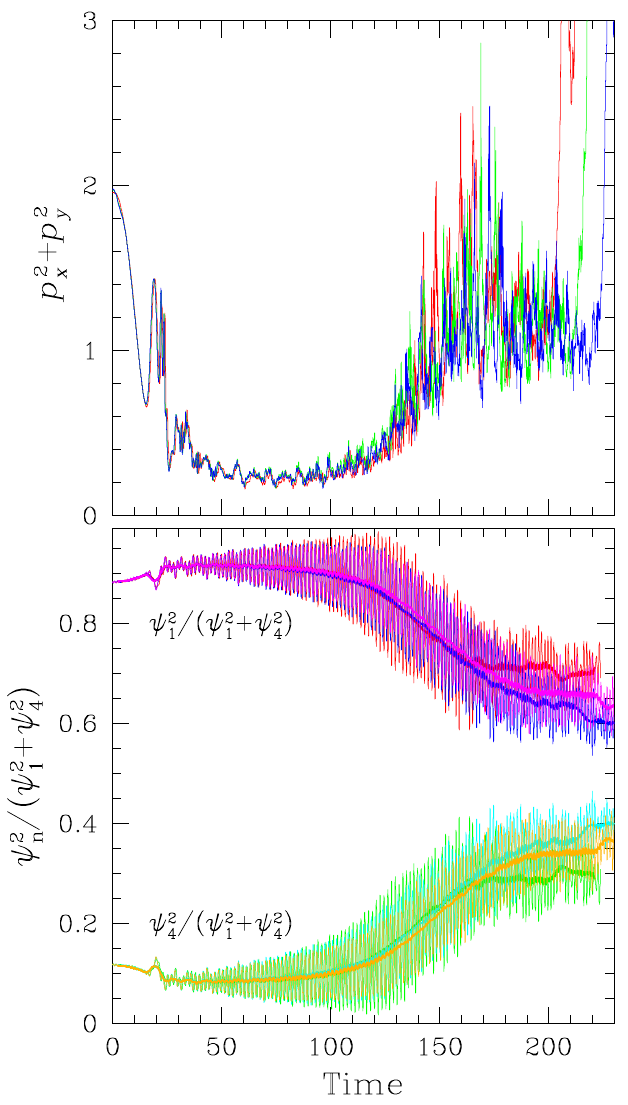}
\caption{ 
Plots of the spatially integrated (linear) momentum squared as a function of time (top), for $l=1,2,3$ in
red, green, and blue, respectively.
The bottom plot shows the instantaneous and time averaged relative values of  %to the fermion number density 
of  $\psi_1^2/(\psi_1^2+\psi_4^2)$ for levels $l=1,2,3$ in red, blue, and magenta, respectively, and
of  $\psi_4^2/(\psi_1^2+\psi_4^2)$ for levels $l=1,2,3$ in green, cyan, and orange, respectively.
The simulations were conducted on $N\times N$ grids where $N=200\times 2^l +1$.
} 
\label{fig:RelPsiConv.pdf}
\end{center}
\end{figure}
Figure \ref{fig:RAS_vs_Psi10.pdf} displays results of scattering angle as a function of boost velocity for a 
range of fermion field strengths.
In the limit of weak fermion field strength, one sees nearly right-angle scattering, consistent with previous results.
However, even in the weak-field limit, electromagnetic forces result in a slight deviation in the vortex trajectory, 
resulting in a small effective impact parameter that precludes perfect right-angle scattering.
As the field strength increases, both sharpness of the transition (with respect to boost velocity)  and the 
value at which the transition occurs decreases.
For low velocity collisions, the vortices are observed to elastically back-scatter similar to traditional vortices.
For high velocity collisions, the vortices undergo deep inelastic scattering and new phenomenology is observed.

Figure \ref{fig:headon_bp} demonstrates high-velocity ($v_b=0.75$) scattering in the strong (fermion) field regime
($\psi_{1,0}=0.4$).
The vortices can be seen to scatter at near right angles like uncharged vortices, but one also sees the production of 
new localized bound states.  These bound states are not vortices as they do not have a topological charge; the 
winding number of the complex scalar field is zero.  The solutions are ``bubble"-like states where the scalar field
interpolates between  the spontaneously broken vacuum of the bulk, $\phi_0$, and the dynamically induced vacuum 
created by the boson-fermion interaction (\ref{eqn:VphiSSB}) that becomes significant when
$\bar{\psi}\psi\rightarrow\kappa_m^{-1}$.
Figure \ref{fig:headon_bp2} displays similar vortex dynamics for a slightly higher fermion 
field strength ($\psi_{1,0}=0.5$). 
The vortices still undergo nearly right angle scattering, but significantly more  interaction can be observed;
 four bubbles are produced and ejected from the collision, while one large bubble remains in the system
 center of mass.
As one continues to increase the fermion field strength of the scattering vortices, one reaches a point at which 
right angle scattering breaks down.    
Figure \ref{fig:headon_bp3} displays vortex dynamics for $\psi_{1,0}=1.5$.  The vortices are observed to 
coalesce for a period of time, creating a pseudo-stable $m=2$ vortex that eventually decays into two 
stable $m=1$ vortices while producing bubbles in the process.  Two stable bubbles can be seen 
ejected orthogonal to the original collision axis while one bubble is again seen remaining in the system center
of mass.  
Unstable bubbles can also be seen to appear  (most clearly around $n=1720$) but eventually disperse, seemingly 
when the surface tension of the scalar bubble outweighs the outward repulsive force from the fermion bound state.

The final example of collisions to be discussed here can be seen in 
Figure \ref{fig:headon_bp4}, which  shows time-elapsed pictures of a collision  with velocity slightly above 
where  the vortices are observed to  back-scatter.  
The vortices are observed to touch, coalesce, interact, and eventually disperse. 
When the scalar vortices remain in close proximity, they create an effective trap 
for the fermion bound states that is significantly larger than the individual harmonic
traps that originally confined each of the    stationary vortices.
The fermion bound states are observed to  stably interact in the larger trap for 
a time $T \approx 100$, and then the dynamics become %turbulent and 
unstable and the fermion bound states 
(and the scalar vortices that form the trap they are contained within) eventually disperse.
To better understand the transition from stable rotation to instability, %turbulence, 
it is helpful to consider the relative contributions of $\psi_1^2$ and $\psi_4^2$
%
%
%to the overall fermion number density
over time
and the implications to linear momentum, orbital angular 
momentum, and spin (Figure \ref{fig:RelPsiConv.pdf}).
In the Dirac representation, the four component spinor representing spin-up particles at rest 
is given by $\psi \propto (1,0,0,0)$.   
In a frame that is boosted in the $x-y$ plane, the new spinor is given by $\psi \propto (1,0,0, (p_x + i p_y)/(E+m))$. 
As such, for a given fermion field strength ($\psi^\dag\psi = \psi_1^2+\psi_4^2$) the relative increase in $\psi_4^2$ can be interpreted as an increase in the 
the average momentum $|p|^2 = p_x^2 + p_y^2$.
For $T\gtrsim 100$ the momentum in the fermion field continues to increase %, while the total energy and particle number remain conserved,
until the vortices begin to radiate energy and eventually disperse.
It is also  helpful to remember that 
in isolation, prior to the interaction of the collision, each fermion bound state is an  eigenstate of total angular momentum
\begin{eqnarray}
\hat{J}_z  
&=& \hat{L}_z + \hat{S}_z \\
&=& -i\partial_\theta + \hat{\Sigma}_z,
\end{eqnarray}
where  $\Sigma_z=\frac{1}{2} {\rm diag}(1,-1,1,-1)$ is the spin projection matrix
and all $\psi$  are eigenstates with  eigenvalue $j_z=+\frac{1}{2}$.    
%
%
%
%
%
%
%which
%
While  interacting, however, since total angular momentum is conserved while  the  intrinsic spin  
$\psi_1^2 - \psi_4^2$ decreases,
Figure \ref{fig:RelPsiConv.pdf} also implies an increase in orbital angular momentum during the mixing of the fermion states. 
This increase in both linear momentum and orbital angular momentum appears to contribute to the eventual instability 
and dispersal of the vortices.

Lastly, it is noteworthy to acknowledge the numerical challenges posed by this low-energy, deeply inelastic scattering class of solutions.
The momentum gained by the fermion field  increases the wavevector and creates gradients in the field that  increase the solution error.
A helpful diagnostic to monitor is  the relative error in the total conserved energy,  
\begin{eqnarray}
E_{\rm error}(t) &=& \frac{E(t) + E_{\rm rad} - E_0 }{E_0},
\label{eqn:EnergyError}
\end{eqnarray}
where $E(t)$ is  current energy in the grid at time $t$,  $E_{\rm rad}$ is the  radiated energy calculated by integrating the
power through  the outer boundary surface from $t=0$ to $t$, and $E_0$ is the energy in the grid at $t=0$.
For most evolutions discussed in this section, $E_{\rm error}(t)$ remained less than or on the order of $10^{-4}$.
However, for the evolution described in Figure \ref{fig:headon_bp4} 
where the fermion field underwent a significant acceleration and spatial contraction, 
the error became significantly higher at late stages of the evolution;
%
%For the evolution displayed in Figure \ref{fig:headon_bp4}, 
the error  
was roughly $10^{-4}$ at $t=150$, but as the momentum (wavevector) of the field increased, the solution error increased
to almost $10^{-2}$ by the end of the evolution when dissipative effects required for stability began to have a
significant effect (Appendix \ref{app:NumericalMethods}).

{\color{blue}
Throughout  this work, %the error described by  (\ref{eqn:EnergyError}) 
$E_{\rm error}(t)$
and the conservation of $\psi^\dag\psi$
were monitored and grid resolutions were used that kept the error within acceptable tolerances.
The methods employed maintain code stability, are convergent, and  disappear in the limit that the lattice spacing goes to zero.  
However, if in future work one desires to study solutions that   dynamically increase in wavevector  for longer periods of time, 
one  may want to employ adaptive mesh refinement or other techniques to more efficiently handle fine resolution features.
}
For additional insight and comparison, 
Appendix \ref{app:SelfInteraction} discusses a toy model of a self-interacting fermion pulse that demonstrates similar underlying phenomenology 
in a simpler example. %with more transparency.

%%%%%%%%%%%%%%%%%%%%%%%%%%%%%%%%%%%%%%%%%%%%%%%%%%%%%
%%%%%%%%%%%%%%%%%%%%%%%%%%%%%%%%%%%%%%%%%%%%%%%%%%%%%
\section{Conclusions}
%%%%%%%%%%%%%%%%%%%%%%%%%%%%%%%%%%%%%%%%%%%%%%%%%%%%%
%%%%%%%%%%%%%%%%%%%%%%%%%%%%%%%%%%%%%%%%%%%%%%%%%%%%%

Results have been presented from numerical simulations of the flat-space nonlinear Maxwell-Klein-Gordon-Dirac 
equations with a  spontaneously broken symmetry and repulsive boson-fermion interaction.   
The findings build upon historical vortex  research
\cite{Nielsen_197345_Original,MBHindmarsh_1995,deVega_PhysRevD.18.2932,deVega_ClassicalVortexSolution,Kleidis_ChargedCosmicStrings,Gleiser_PhysRevD.76.041701,SHELLARD_1988262,RUBACK_1988669,MORIARTY_1988411,Myers_PhysRevD.45.1355,Dziarmaga_PhysRevD.49.5609,Abbott_PhysRevD.97.102002,Helfer_PhysRevD.99.104028,Pillado_PhysRevD.100.023535}
and complement other work that has explored fermionic bound states within vortices %with other types of interactions 
\cite{Nohl_PhysRevD.12.1840,Jackiw_1981681_ZeroModes,Kleidis_ChargedCosmicStrings,Lozano_PhysRevD.38.601} 
by including massive Dirac fermions similar to those explored widely in the context of
condensed matter physics 
\cite{Krasnov_BFHModelHilbertSpace,Bukov_PhysRevB.89.094502,Milczewski_PhysRevA.105.013317,Greiner_OpticalLattice1,Greiner2_PhysRevLett.87.160405,Fehrmann_200423,Cramer_PhysRevLett.93.190405,Albus_PhysRevA.68.023606,Lewenstein_PhysRevLett.92.050401,Wehling_02012014,Classen_PhysRevB.93.125119,Ye_MassiveDiracKagome,Yang_MassiveDirac,Lin_PhysRevB.102.155103}.

Massive fermion bound state solutions have been shown to exist in the cores of scalar vortices when a repulsive boson-fermion interaction 
is present.
Closed-form solutions were obtained that determined the dispersion relation and  relationship between 
the interaction strength $\kappa_m$, effective fermion mass $\kappa_d^{-1}$, and width of bound state $\sigma$,
for approximate solutions when sufficiently  contained within the scalar vortex and when the fermion
field strength  did not significantly perturb the spontaneously broken vacuum of the condensate.
{\color{blue}
The effective mass $\kappa_d^{-1} \propto m_\F$ determines the mass gap, while the interaction strength $\kappa_m$
determines the energy band of allowed states.
When considered in the context of typical bulk parameters, the predicted mass gap and energies of bound states were shown to be 
comparable to those observed in gapped Dirac materials
\cite{KUMAR_2014S136,Fischer_ArmchairGraphene}.
}
Numerical solutions to the stationary equations of motion were then obtained that confirmed the existence of bound states
and demonstrated where the closed form approximations ceased to be valid.
The stationary bound state solutions were then time-evolved and demonstrated  stability until 
the the fermion field was strong enough to significantly change the vacuum expectation value of the condensate,
$\bar{\psi}\psi\rightarrow \kappa_m^{-1}$.

Simulations of head-on collisions of  fermion bound states in vortices were conducted for a range of fermion 
field strengths  and boost  velocities. 
For low fermion field strength, head-on scattering closely resembled traditional right angle scattering 
\cite{SHELLARD_1988262,RUBACK_1988669,MORIARTY_1988411}, but
solutions still deviated from perfect right angle scattering due to  Coulombic interactions between  
the charged  fermion field and the magnetic field of the scalar vortex, 
resulting in the impact parameter dynamically moving away from zero.  
As one increases fermion field strength, there are distinctly different behaviors for low and high velocity scattering.

For lower velocity deep inelastic scattering, charged vortices were observed to coalesce, interact, become unstable, %turbulent, 
and disperse.  The instability %turbulence 
arises when the fermion bound states are confined within the large trap formed by the close
proximity of the two scalar vortices. The larger  trap  is flatter and no longer spatially harmonic, and
there is a dynamic mix of time-varying  radial and axial electromagnetic fields that drive the dynamics of the fermion field.
The mixing of bound states drives  a decrease in total spin, which demands an
increase in orbital angular momentum (in order to  conserve total angular momentum), 
leading to  instability and eventual dispersal.
Additional understanding of the fermion field dynamics as instability %turbulence 
arises could be a productive 
area for future study.

For high velocity scattering and low  fermion field strength, 
the vortices underwent near-right-angle scattering, and the creation of a new type of  
nontopological  scalar-fermion bound state  was observed.
The bound states are bubble-like in that the scalar field interpolates between 
the spontaneously broken vacuum of the bulk, $\phi_0$,  and the  vacuum of  (\ref{eqn:VphiSSB}) 
that becomes dynamically shifted away from $\phi_0$ for large $\bar{\psi}\psi$. 
The solutions are reminiscent of the oscillons created by the vortex-antivortex collisions
of  Gleiser and Thorarinson
\cite{Gleiser_PhysRevD.76.041701}, but the scalar bound states observed here 
do not have a harmonic time dependence and are therefore not oscillons;
they arise from a balance between the outward  self-repulsion of the fermion field 
(that does have a harmonic time dependence and nonzero orbital angular momentum) 
and the inward surface tension of the scalar bubble. 
The solutions  also bring to mind a time-reversed version of the solutions  of 
Srivastava \cite{Srivastava_PhysRevD.46.1353_BUBBLES},  who discussed the creation of vortices 
by the interaction of both critical and subcritical bubbles in the context of cosmological phase transitions.

The realization of this work 
that  vortex-vortex interactions can  lead to the production of boson-fermion bubbles 
complements  the findings of \cite{Srivastava_PhysRevD.46.1353_BUBBLES} and similiarly
could  have impact on the understanding of gauged Abelian-Higgs phase transitions 
in models with a boson-fermion interaction.
While the consequences of the boson-fermion interaction introduced here to the electronic structure of the bulk material 
were not discussed, this work provides insight into the existence, stability, and dynamics of bound states that could
arise in condensed matter systems that describe  quasiparticles in Bose-Fermi models similar to the model (\ref{eqn:OverallLagrangian}). 
Finally, a dedicated and more detailed analysis %of the attributes 
of the boson-fermion bubbles  discovered here, 
particularly their stability, may also be an interesting area
for future study.
%

%

%%%%%%%%%%%%%%%%%%%%%%%%%%%%%%%%%%%%%%%%%%%%%%%%%%%%%%%
%%%%%%%%%%%%%%%%%%%%%%%%%%%%%%%%%%%%%%%%%%%%%%%%%%%%%%%
%%%%%%%%%%%%%%%%%%%%%%%%%%%%%%%%%%%%%%%%%%%%%%%%%%%%%%%
%%%%%%%%%%%%%%%%%%%%%%%%%%%%%%%%%%%%%%%%%%%%%%%%%%%%%%%
%%%%%%%%%%%%%%%%%%%%%%%%%%%%%%%%%%%%%%%%%%%%%%%%%%%%%%%
\appendix
%%%%%%%%%%%%%%%%%%%%%%%%%%%%%%%%%%%%%%%%%%%%%%%%%%%%%%%
%%%%%%%%%%%%%%%%%%%%%%%%%%%%%%%%%%%%%%%%%%%%%%%%%%%%%%%
%%%%%%%%%%%%%%%%%%%%%%%%%%%%%%%%%%%%%%%%%%%%%%%%%%%%%%%
%%%%%%%%%%%%%%%%%%%%%%%%%%%%%%%%%%%%%%%%%%%%%%%%%%%%%%%
%%%%%%%%%%%%%%%%%%%%%%%%%%%%%%%%%%%%%%%%%%%%%%%%%%%%%%%

%%%%%%%%%%%%%%%%%%%%%%%%%%%%%%%%%%%%%%%%%%%%%%%%%%%%%%%
%\section{Numerical Methods \label{app:NumericalMethods}}
\section{\uppercase{Numerical Methods} \label{app:NumericalMethods}}
%%%%%%%%%%%%%%%%%%%%%%%%%%%%%%%%%%%%%%%%%%%%%%%%%%%%%%%

%%%%%%%%%%%%%%%%%%%%%%%%
\begin{figure}
\begin{center}
\includegraphics[width=80mm]{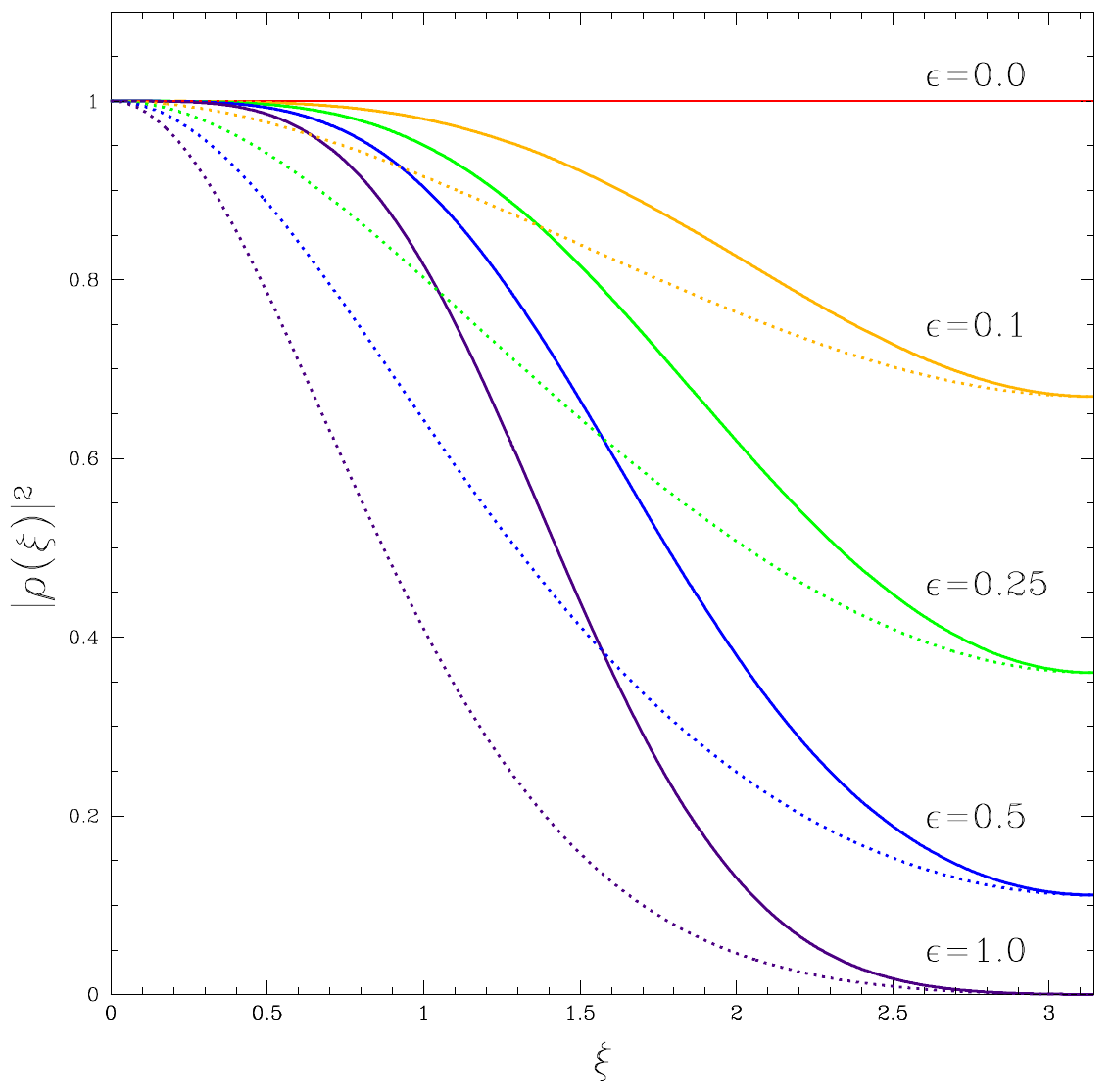}
\caption{
Plots of amplification factors of modes for a physically motivated Wilson term (dotted lines) as it compares
to a numerically motivated Kreiss-Oliger dissipation approach (solid lines).  Plots for $\epsilon=\{0, 0.1, 0.25, 0.5, 1.0\}$
are shown in red, orange, green, blue and indigo, respectively.  
%The Kreiss-Oliger amplification factor is plotted with solid lines
%and the amplification factor with a Wilson term in dotted lines. 
For a given $\epsilon$, the Kreiss-Oliger approach has equal suppression at $\xi=\pi$ while having less attenuation for lower
wavevectors. } 
\label{fig:AmpFacs.pdf}
\end{center}
\end{figure}
%%%%%%%%%%%%%%%%%%%%%%%%

This appendix describes the numerical methods for addressing potential numerical instabilities  
(including fermion doubler modes), 
the  time evolution and boosting of charged vortices, 
and a number of self-consistency and monitoring techniques to ensure solution accuracy.

{
\color{blue}
The time-evolution methods employed throughout  this work employ second-order finite difference 
approximations to continuous (not lattice) equations of motion.
While all finite difference approximations of partial differential equations may be susceptible to numerical
instabilities, the modeling of  Dirac fields using discretized approximations can be particularly challenging.
Adding a ``Wilson term" \cite{Wilson_PhysRevD.10.2445,deResende_PhysRevB.96.161113}
to the Dirac equations of motion is a common approach to  directly address %so-called
\emph{fermion doublers} on a lattice; the approach focuses on suppressing
the $k\approx \pi/a$ mode for a lattice spacing $a$, which %in numerical terms 
is  the Nyquist limit of the numerical discretization.
The approach is physically motivated and introduces an ultraviolet cutoff (low-pass filter) by adding a term 
proportional to $\partial^2_\mu\psi$ to the equations of motion to specifically address the doubler modes.
Figure \ref{fig:AmpFacs.pdf} compares the result of using a Wilson term in the equations of motion
to the numerically motivated approach of Kreiss-Oliger dissipation (the technique employed in this work).
Kreiss-Oliger dissipation similarly adds a higher order derivative term ($\partial^4_\mu\psi$) that also
suppresses high-wavevector modes. 
Using Von Nuemann stability analysis yields the following amplification factors for second order
Crank-Nicolson time-centered difference schemes for the addition of a Wilson term and 
Kreiss-Oliger dissipation, respectively,

\begin{eqnarray}
\left|\rho(\xi)\right|^2_{\rm W} 
&=& 
\frac{\ds 1 + i\frac{\lambda}{2} \sin(\xi) - \frac{\epsilon}{2} \left(\cos(\xi)-1\right)}{\ds 1 - i\frac{\lambda}{2} \sin(\xi) + \frac{\epsilon}{2} \left(\cos(\xi)-1\right)},
\label{eqn:AmpFacW}
\\
\left|\rho(\xi)\right|^2_{\rm KO} &=& 
\frac{\ds 1 + i\frac{\lambda}{2} \sin(\xi) - \frac{\epsilon}{8} \left(3 - 4\cos(\xi) + \cos(2\xi)\right)}
{\ds 1 - i\frac{\lambda}{2} \sin(\xi) +\frac{\epsilon}{8} \left(3 - 4\cos(\xi) + \cos(2\xi)\right)},
\label{eqn:AmpFacKO}
\nonumber \\
\end{eqnarray}
where $\xi=k a$ for lattice spacing $a$.
These amplification factors can be seen in  Figure \ref{fig:AmpFacs.pdf}  for multiple values of the tunable parameter $\epsilon$.
In the Wilson term, one typically uses $\epsilon = r a$ %for the lattice space $a$ 
with a tunable parameter $r$.  
The Courant factor $\lambda=dt/dx$ was set to $\lambda = 0.5$ in all simulations used  in this work and in Figure \ref{fig:AmpFacs.pdf}. 
%(\ref{eqn:AmpFacW}) and (\ref{eqn:AmpFacKO}).
%
Both approaches have tunable scaling parameters and vanish in the $a\rightarrow 0$ limit.  
While the Wilson term may be more appealing from a physical (vice numerical) perspective, 
the Kreiss-Oliger approach 
is a higher order technique with  a sharper frequency response that suppresses the doubler mode 
equally while having less negative effect on low-frequency modes.
%
%
%An important aspect to emphasize is that if high-wavevector  components emerge over time, these 
%techniques will suppress the growth of such modes in a dissipative and non-conservative way that
%would be observed by lack of conservation of energy or 
%$\psi^\dag\psi$.
%
%
%Furthermore, the solution error is monitored and  evolutions are terminated or
%performed again at higher resolution to stay within a desired error tolerance.
%
%
Most importantly, if doubler modes or high-wavevector instabilities  become prevalent in a solution, 
the dissipative approach employed here leads to 
solution errors that  are  quantified with (\ref{eqn:EnergyError}) and can be monitored.
}

The following equations describe the transformation of the scalar, fermion, and electromagnetic 
fields from at-rest  solutions in cylindrical coordinates ($R,\theta$)
%to at-rest solutions in Cartesian coordinates, 
to boosted solutions in Cartesian coordintes ($\tilde{t},\tilde{x},\tilde{y})$;
the boosted scalar and electromagnetic fields are given by
\begin{eqnarray}
%%%%%%%%%%
\phi_1
(\tilde{t},\tilde{x},\tilde{y})
&=&   
\phi(R)\cos\theta, 
\label{eqn:FirstBoost}\\
%%%%%%%%%%
\phi_2 
(\tilde{t},\tilde{x},\tilde{y})
&=& \phi(R)\sin\theta,  \\
%%%%%%%%%%
\Pi_1 
(\tilde{t},\tilde{x},\tilde{y})
&\approx& 
\frac{\phi_1\left( \tilde{t}+\Delta\tilde{t},\tilde{x},\tilde{y}\right)-\phi_1\left( \tilde{t}-\Delta\tilde{t},\tilde{x},\tilde{y}\right)}{2 \Delta\tilde{t}},
\hspace{5mm}
\\
%%%%%%%%
\Pi_2
(\tilde{t},\tilde{x},\tilde{y})
 &\approx& 
\frac{\phi_2\left( \tilde{t}+\Delta\tilde{t},\tilde{x},\tilde{y}\right)-\phi_2\left( \tilde{t}-\Delta\tilde{t},\tilde{x},\tilde{y}\right)}{2 \Delta\tilde{t}}, 
\\
%\end{eqnarray}
%\begin{eqnarray}
%%%%%%%%%%
A_{\tilde{t}} 
(\tilde{t},\tilde{x},\tilde{y})
%%%%
&=& 
\gamma  \left( A_t(R) -v_0\tilde{A}_\theta(R) \sin\theta \right), 
\\
%%%%%%%%%%%%%
A_{\tilde{x}} 
(\tilde{t},\tilde{x},\tilde{y})
%%%%%%
&=& 
\gamma \left( v_0 A_t(R)
-\tilde{A}_\theta(R) \sin\theta \right),
\\
%%%%%%%%%%%%%
A_{\tilde{y}} 
(\tilde{t},\tilde{x},\tilde{y})
&=& 
 \tilde{A}_\theta(R) \cos\theta,
\\
%%%%%%%%%%
E_{\tilde{x}}
(\tilde{t},\tilde{x},\tilde{y})
&=& 
  E_R(R) \cos\theta, 
\\
%%%%%%%%%%
E_{\tilde{y}}
(\tilde{t},\tilde{x},\tilde{y})
&=& 
\gamma  E_R(R) \sin\theta 
- \gamma v_0B_z(R),
\\
%%%%%%%%%%
B_{\tilde{z}}
(\tilde{t},\tilde{x},\tilde{y})
&=& 
 \gamma B_z(R)
-\gamma v_0  E_R(R)  \sin\theta,
\\
%%%%%%%%%%%%%
%%%%%%%%%%
E_{\tilde{z}}(\tilde{t},\tilde{x},\tilde{y}) &=& 
B_{\tilde{x}}(\tilde{t},\tilde{x},\tilde{y}) =
B_{\tilde{y}}(\tilde{t},\tilde{x},\tilde{y})=0,
\text{ and} \\
 A_{\tilde{z}} 
(\tilde{t},\tilde{x},\tilde{y})
&=& 0,
\end{eqnarray}
while the boosted fermion field components are given by
\begin{eqnarray}
%%%%%%%%%%%%%%%%%%%%%%%%%%%%%%%%%
a_1
(\tilde{t},\tilde{x},\tilde{y})
%%%%%%%%
&=&
\cosh\left(\lambda/2\right) \Psi_1\left( R\right) 
\cos\left({\omega  \gamma\left( \tilde{t} + v_0 \tilde{x}\right)} \right)
\nonumber \\ &&
\pm \sinh\left(\lambda/2\right)\Psi_4\left( R  \right)
\sin\left( \theta - \omega  \gamma\left( \tilde{t} + v_0 \tilde{x}\right) \right),
%e^{i\phi} e^{-i\omega  \gamma\left( \tilde{t} + v_0 \tilde{x}\right)} 
\nonumber \\ \\
%%%%%%%%%%%%%%%%%%%%%%%%%%%%%%%%%
a_2
(\tilde{t},\tilde{x},\tilde{y})
%%%%%%%%
&=&
-\cosh\left(\lambda/2\right) \Psi_1\left( R\right) 
\sin\left({\omega  \gamma\left( \tilde{t} + v_0 \tilde{x}\right)} \right)
\nonumber \\ &&
\mp \sinh\left(\lambda/2\right)\Psi_4\left( R  \right)
\cos\left( \theta - \omega  \gamma\left( \tilde{t} + v_0 \tilde{x}\right) \right),
%e^{i\phi} e^{-i\omega  \gamma\left( \tilde{t} + v_0 \tilde{x}\right)} 
\nonumber \\ 
\\
%\end{eqnarray}
%%%%%%%%%%%%%%%%%%%%%%%%%%%%%%%%%%
%%%%%%%%%%%%%%%%%%%%%%%%%%%%%%%%%%
%\begin{eqnarray}
b_1 
(\tilde{t},\tilde{x},\tilde{y})
&=& 
\mp \cosh\left(\lambda/2\right) 
\Psi_4\left( R \right)
\sin\left(\theta - \omega  \gamma\left( \tilde{t} + v_0 \tilde{x}\right) \right)
\nonumber \\ &&
- \sinh\left(\lambda/2\right)
\Psi_1\left(  R\right) 
\cos\left(\omega  \gamma\left( \tilde{t} + v_0 \tilde{x}\right)\right), \text{ and}
%e^{-i\omega  \gamma\left( \tilde{t} + v_0 \tilde{x}\right)}
\nonumber \\ 
 \\
%%%%%%%%
b_2 
(\tilde{t},\tilde{x},\tilde{y})
&=& 
\pm \cosh\left(\lambda/2\right) 
\Psi_4\left( R \right)
\cos\left(\theta - \omega  \gamma\left( \tilde{t} + v_0 \tilde{x}\right) \right)
\nonumber \\ &&
+ \sinh\left(\lambda/2\right)
\Psi_1\left(  R\right) 
\sin\left( \omega  \gamma\left( \tilde{t} + v_0 \tilde{x}\right)\right),
\nonumber \\ 
\label{eqn:LastBoost}
%%%%%%%%%%
\end{eqnarray}
where
\begin{eqnarray}
\lambda &=& \tanh^{-1}(v_0), \\
R &=& \sqrt{\gamma^2\left( \tilde{x} + v_0 \tilde{t}\right)^2 + \tilde{y}^2}, \\
%\sin\phi &=& \frac{\tilde{y}}{R} \\
%\cos\phi &=&\frac{ \gamma\left( \tilde{x} + v_0 \tilde{t}\right) }{R}  
\sin\theta &=& \tilde{y} /R, \\
\cos\theta &=& \gamma\left( \tilde{x} + v_0 \tilde{t}\right) / R, \text{ and}\\
\theta &=& \tan^{-1}\left( \frac{\tilde{y}}{\gamma\left(\tilde{x}+v_0\tilde{t}\right)} \right).
\end{eqnarray}
\begin{figure}[]
\begin{center}
\includegraphics[width=85mm]{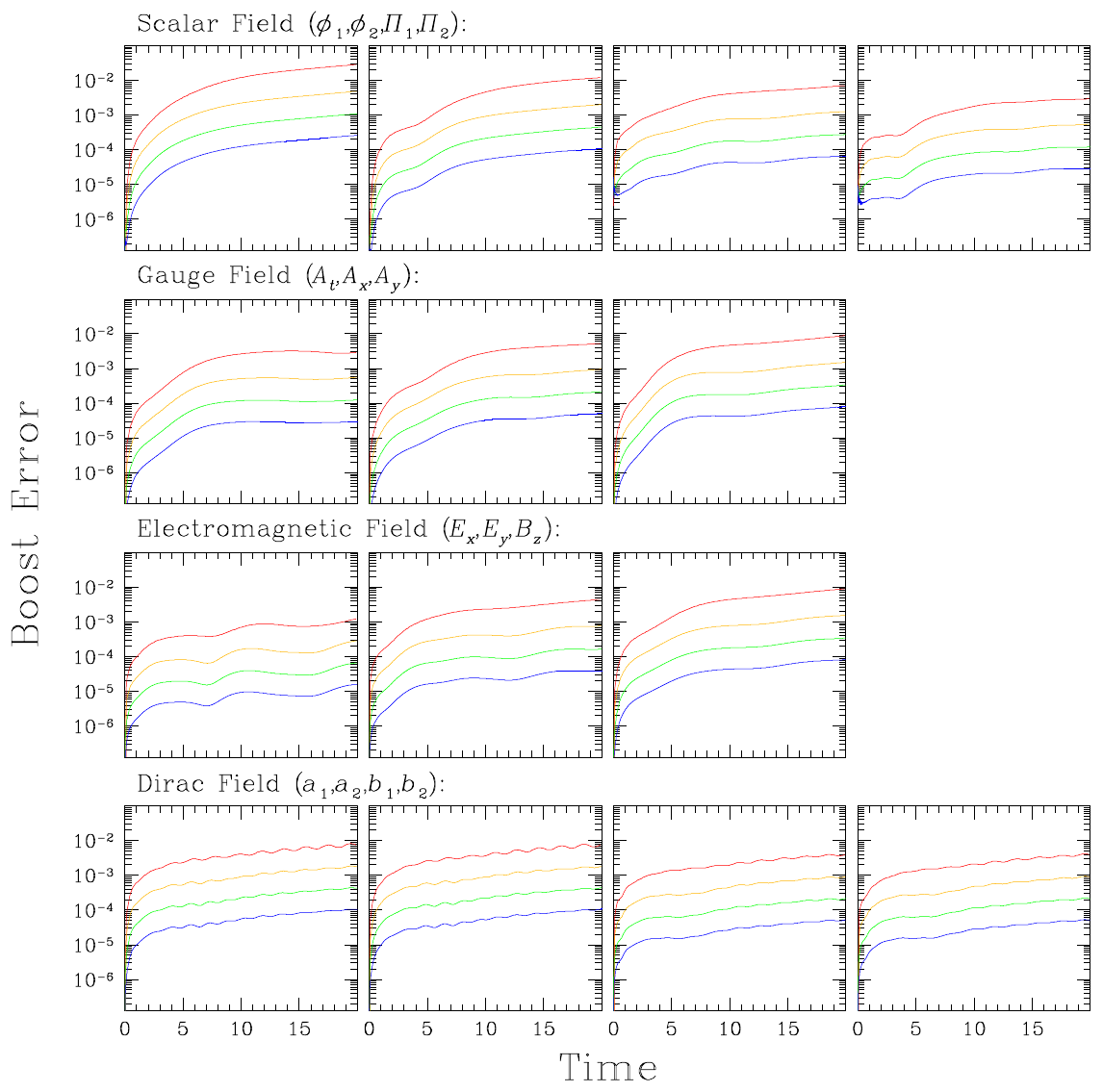}
\caption{ 
Plots of the boost error over time for all  fields when a single vortex is
boosted with $v_x=0.5$.
The error is obtained by calculating the L2-norm of the difference between the 
time-evolved boosted initial data 
and
the analytically boosted  stationary solutions.
Each field is evolved on an $N\times N$ grid where $N= 200\times 2^l + 1$ for levels $l=0, 1, 2, 3$ in red, orange, green,
and blue, respectively,
 with domain spanning 
$\{x:-20<x<20\}$ and 
$\{y:-20<y<20\}$.
The error in all fields is observed to converge quadratically to zero.
} 
\label{fig:ConvergenceBoost.pdf}
\end{center}
\end{figure}
\begin{figure}[]
\begin{center}
\includegraphics[width=80mm]{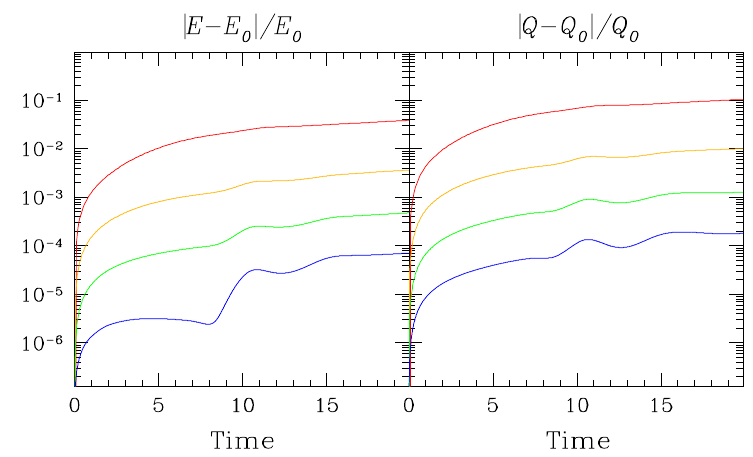}
\caption{ 
Plots demonstrating conservation of energy (left) and charge (right) for the time evolution of 
the (2+1) equations of motion for a head-on collision of two vortices.  The relative change 
in energy (charge) is shown for levels $l=0,1,2,3$ in red, orange, green, and blue, respectively, 
for an $N\times N$ grid where $N= 200\times 2^l + 1$.
The error in all fields is observed to converge quadratically to zero. } 
\label{fig:EnergyCharge.pdf}
\end{center}
\end{figure}
Equations (\ref{eqn:FirstBoost}-\ref{eqn:LastBoost}) were used in Section \ref{sec:Scattering} to generate 
initial data for head-on (zero impact parameter) collisions
by setting $\tilde{t}=0$ and superimposing  solutions of the stationary equations 
of motion that were translated to $(\tilde{x},\tilde{y})=(\pm5,0)$ with boost velocities $\mp v_0$
and then time-evolved using (\ref{eqn:First2p1Evol}-\ref{eqn:Last2p1Evol}).
In this section, 
a similar approach is used,
but with a \emph{single} vortex boosted in the 
positive $x$-direction.  
Those time-evolved solutions are then compared to directly transforming a stationary charged vortex solution 
of (\ref{eqn:FirstStataionary}-\ref{eqn:LastStataionary})  with 
(\ref{eqn:FirstBoost}-\ref{eqn:LastBoost}) 
to obtain  boosted solutions for an arbitrary $\tilde{t}$.
Figure \ref{fig:ConvergenceBoost.pdf} shows plots for each field of the boost error, which is defined to be the 
L2-norm of the difference between these two approaches and serves as a very strong method of verification 
of the boost methodology.
It can be seen that every field component has boost error that converges to zero in accordance with the 
second order finite difference scheme being employed.

\begin{figure}[]
\begin{center}
\includegraphics[width=80mm]{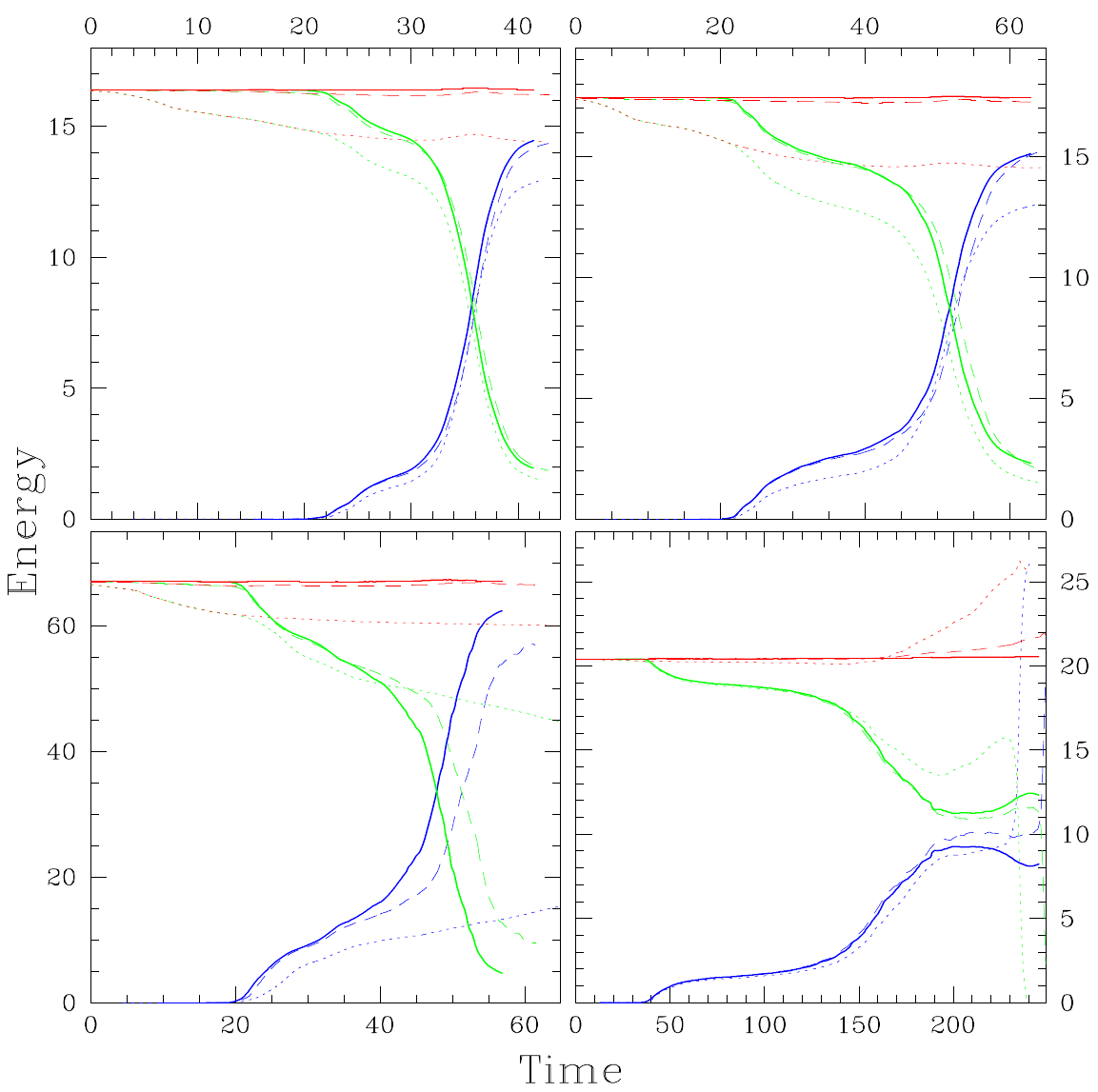}
\caption{ 
Plots of total energy (red), energy currently in the computational domain (green), and cumulative radiated energy (blue).
The upper left figure corresponds to the evolution from Figure \ref{fig:headon_bp};
the upper right figure corresponds to the evolution from Figure \ref{fig:headon_bp2};
the lower left figure corresponds to the evolution from Figure \ref{fig:headon_bp3};
and the lower right figure corresponds to the evolution from Figure \ref{fig:headon_bp4}.
Simulations were performed on an $N\times N$ grid where $N=200\times 2^l + 1$ for $l=1,2,3$,
with dotted, dashed, and solid lines, respectively.
Total energy is observed to be conserved in a convergent manner. 
}
\label{fig:EnergyOverTime.pdf}
\end{center}
\end{figure}

Next, Figure \ref{fig:EnergyCharge.pdf} demonstrates energy and charge conservation of two charged 
vortices boosted at one another that undergo near-right-angle scattering.  The error in both
energy and charge conservation can also be seen to converge to zero with second order accuracy as desired.

Lastly, Figure \ref{fig:EnergyOverTime.pdf} displays the energy conservation for each of the simulations
discussed in Figures \ref{fig:headon_bp}-\ref{fig:headon_bp4}.
For each simulation, the total energy, energy in the grid, and  cumulative radiated energy are  displayed 
for multiple grid resolutions.
The cumulative radiated energy at time $t$ is calculated by integrating the power radiated through a Gaussian surface 
around the computational domain from the beginning of the simulation to  time $t$. The cumulative radiated energy
is then added to the energy in the grid at $t$ to obtain the total energy at $t$.
For the $l=3$ grids, %($1601\times 1601$ gridpoints), 
the simulations displayed by 
Figures \ref{fig:headon_bp}-\ref{fig:headon_bp3} demonstrate total energy conservation to a few parts 
in $10^4$.
{\color{blue}
The simulation displayed by \hbox{Figure \ref{fig:headon_bp4}} demonstrates similar performance until the 
$m=2$ vortex becomes unstable and disperses.   
By the end of the simulation,  conservation of total energy is demonstrated to slightly better than 
one part in $10^2$. 
Whether the higher error in these solutions arises  from doubler modes related to  the discretization of the Dirac equation
or  simply from accelerating fields that increase the wavevector,  the resultant error is monitored and understood.  
}

%%%%%%%%%%%%%%%%%%%%%%%%%%%%%%%%%%%%%%%%%%%%%%%%%%%%%%%
\section{\uppercase{Self-interaction in the 1D Maxwell-Dirac equation}  \label{app:SelfInteraction}}
%%%%%%%%%%%%%%%%%%%%%%%%%%%%%%%%%%%%%%%%%%%%%%%%%%%%%%%

\begin{figure}[]
\begin{center}
\includegraphics[width=80mm]{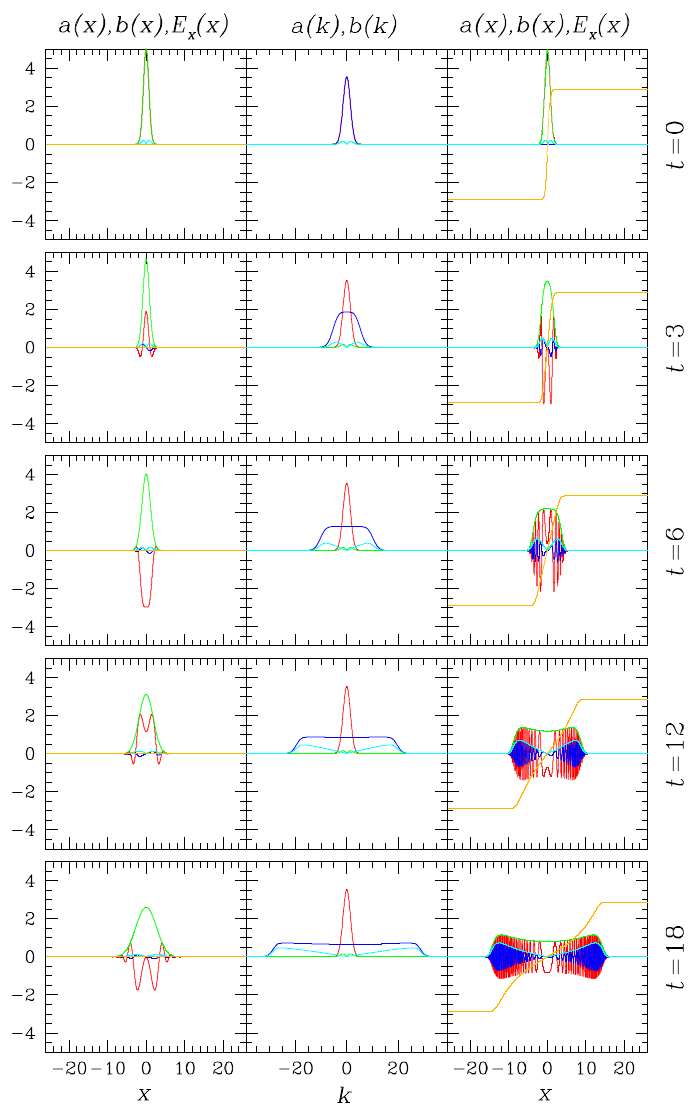}
\caption{ 
Plots of  $a(x)$, $|a(x)|$, $b(x)$, $|b(x)|$, and $E_x(x)$  for the 1D plane-symmetric free fermion field (left column) 
and self-interacting fermion and Maxwell fields (right column), 
and $a(k)$ and $b(k)$ for both 
fields (center column).
The $a(x)$, $|a(x)|$, $b(x)$, $|b(x)|$, and $E_x(x)$ fields are plotted in red, green, blue, cyan, and orange, respectively.
 $a(k)$ and $b(k)$ for the free field are plotted in red and green, while 
$a(k)$ and $b(k)$ for the self-interacting field are in blue and cyan.
The free field is observed to disperse with constant spectral content  $a(k)$ and $b(k)$, while the self-interacting 
field disperses with an accelerating wave front and increasing wavevector over time.
} 
\label{fig:FreeSelf.pdf}
\end{center}
\end{figure}

\begin{figure}[]
\begin{center}
\includegraphics[width=80mm]{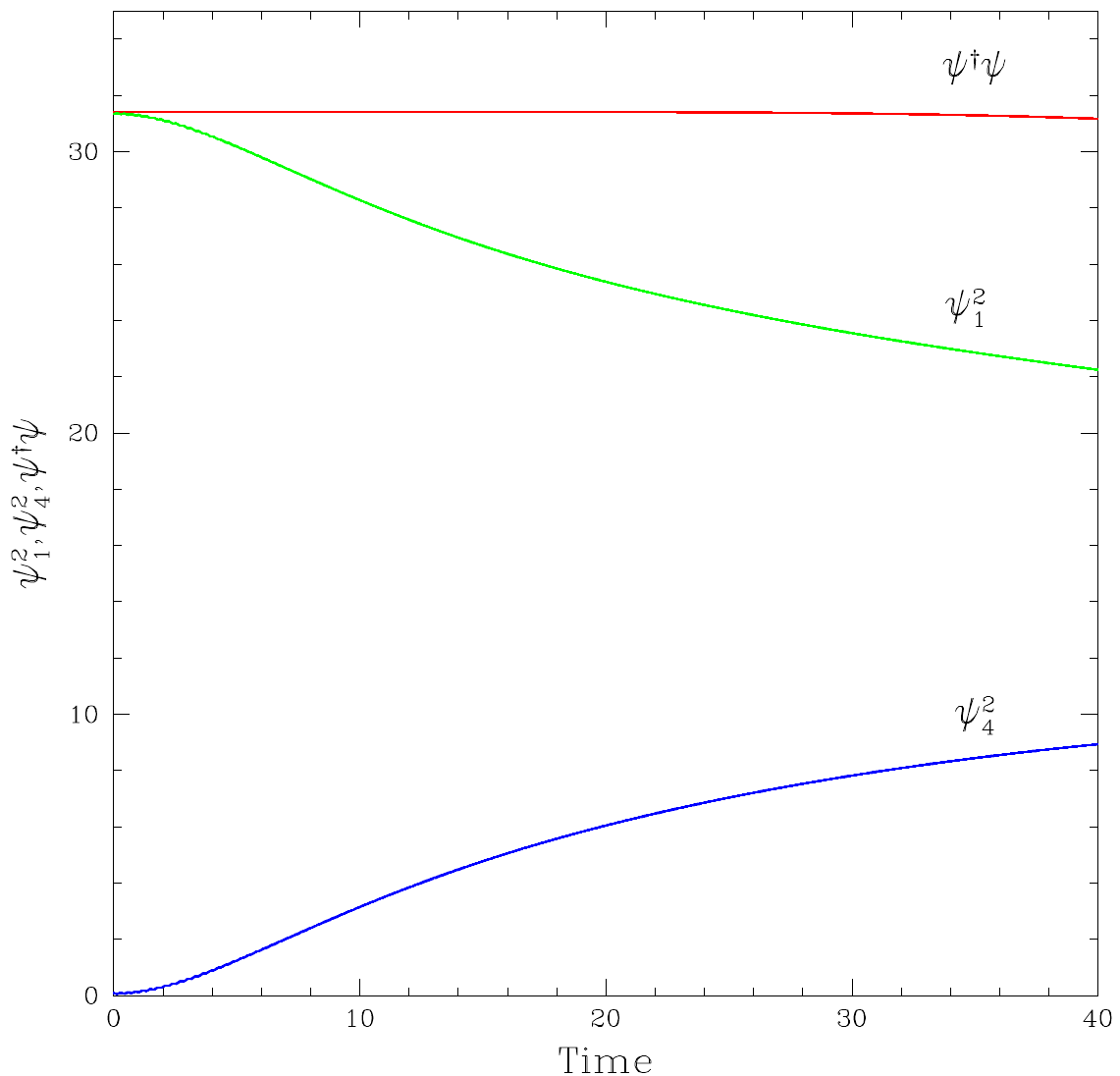}
\caption{ 
Plots of $\psi_1^2$, $\psi_4^2$, and $\psi^\dag\psi$ (red, green, and blue, respectively)
for the  1D plane-symmetric  self-interacting field. 
The total fermion field strength $\psi^\dag\psi$ is conserved, while 
$\psi_1^2$ decreases, and  $\psi_4^2$  (which is proportional to $p_x^2$) increases.
} 
\label{fig:SimplePsi14.pdf}
\end{center}
\end{figure}

While an interesting and related in-depth analysis of the self-interaction properties of the classical and 
quantum Maxwell-Dirac equations can be found in  \cite{Lv_PhysRevA.90.034101}, 
this appendix provides a simple and intuitive example that %directly 
more simply demonstrates
the effects seen by the interacting fermion fields confined to vortices described in this work,
particularly the last scattering example in \hbox{Section \ref{sec:Scattering}} that demonstrated unstable % turbulent 
behavior and is visualized in Figure \ref{fig:headon_bp4}.

A toy model is desired that is as close as possible to (\ref{eqn:OverallLagrangian}) to remain most relevant to this work,
while also removing boson-fermion interactions and focusing solely on the properties of the fermion field.
Setting $\phi=0$ everywhere, the model essentially becomes
\begin{eqnarray}
\mathcal{L} &=& 
\sqrt{-g}\left( L_{M} +
L_{D} 
\right).
\label{eqn:OverallLagrangianSimplified}
\end{eqnarray}
Slightly different dimensionless variables are required since the ones used in the body of this work are  in 
terms of condensate parameters that are now being ignored. 
However, one can use $\xi$ as an arbitrary parameter to set the relative 
length scale of the fermion bound state and use $\kappa_d = \lambdabar/\xi$ as the dimensionless 
model parameter.
This leads to 1D equations of motion, 
\begin{eqnarray}
%%%%%%%%%%%%%%%%%%%%%%%%%%%%%%%%%%%%%%%%%%%%%%%%
%%%%%%%%%%%%%%%%%%%%%%%%%%%%%%%%%%%%%%%%%%%%%%%%
\partial_t a_1 
&=& -\partial_xb_1 - \frac{1}{2} A_ta_2  - \frac{1}{2} A_xb_2  
+ \frac{1}{2} A_yb_1 
+ \kappa_d^{-1}a_2, 
\label{eqn:First1dDirac}
\nonumber \\ \\
%%%%%%%%%%%%%%%%%%%%%%%%%%%%%%%%%%%%%%%%%%%%%%%%
\partial_ta_2 &=& -\partial_xb_2 + \frac{1}{2} A_ta_1  + \frac{1}{2} A_xb_1  
+ \frac{1}{2} A_yb_2
- \kappa_d^{-1}a_1,
\nonumber \\ \\
%
%
%%%%%%%%%%%%%%%%%%%%%%%%%%%%%%%%%%%%%%%%%%%%%%%%
\partial_t b_1 &=& -\partial_xa_1 - \frac{1}{2} A_t b_2 - \frac{1}{2} A_xa_2  
- \frac{1}{2} A_ya_1 
- \kappa_d^{-1}b_2,
\nonumber \\ \\
%%%%%%%%%%%%%%%%%%%%%%%%%%%%%%%%%%%%%%%%%%%%%%%%
\partial_t b_2 &=& -\partial_xa_2 + \frac{1}{2} A_t b_1 + \frac{1}{2} A_xa_1  
- \frac{1}{2} A_ya_2
+  \kappa_d^{-1}b_1,
\nonumber \\ \\
%%%%%%%%%%%%%%%%%%%%%%%%%%%%%%%%%%%%%%%%%%%%%%%%
%%%%%%%%%%%%%%%%%%%%%%%%%%%%%%%%%%%%%%%%%%%%%%%%
\partial_t E^x  
&=&
- \alpha_{\M}^{-1}\kappa_d^{-2}
\left(
a_1b_1 + a_2b_2
\right),
\\
 \partial_t E^y  
&=&
-\partial_xB^z 
-   \alpha_{\M}^{-1}\kappa_d^{-2}
\left(
a_1b_2 - a_2b_1\right),
\\
\partial_t B_z  &=& -\partial_x E_y, 
\\
\partial_tA_x &=&  -E_x + \partial_xA_t, \\
\partial_tA_y &=&  -E_y,  \text{ and}\\
%
%\partial_tA_z &=&  -E_z  \\
%
%
\partial_tA_t &=&  \partial_xA_x,  
\label{eqn:Last1dDirac}
\end{eqnarray}
where $\alpha_\M = 16 \pi \alpha'$ for the fine structure constant $\alpha'$. 
Figure \ref{fig:FreeSelf.pdf} shows time evolution of these equations for 
the self-interacting fermion field and the free fermion field ($A_\mu=0$).
The initial data are taken to be %gaussian for $a_1$,
\begin{eqnarray}
a_1 &=& a_{1,0} \exp\left(-x^2/\sigma^2\right), \\
a_2 &=& 0, \\
b_1 &=& 0, \text{ and} \\
b_2 &=& \kappa_d x a_1, 
\end{eqnarray}
and the appropriate Maxwell equations are used to solve for the electromagnetic fields.
The fields are time-evolved using equations (\ref{eqn:First1dDirac}-\ref{eqn:Last1dDirac}),
and discrete Fourier transforms are taken at each time to obtain $a(k)$ and $b(k)$ for 
both the free and interacting fields. 
The Gaussian initial data, $a_1(t)$, naturally give rise to Gaussian $a(k)$. % at $t=0$.  
For the free field, the solutions are the well-known 1D solutions and disperse in accordance
with the dispersion relation $\omega^2 = \sigma^{-2} + \kappa_d^{-2}$; the field 
disperses in the spatial domain, and the spectral content $a(k)$ and $b(k)$ does not change over time.
 For the self-interacting field, however, the simple dispersion relation does not hold and the wave front of the
 dispersing fermion field accelerates based on strong Coulombic self-repulsion.
As the field accelerates, $p_x$ increases, which increases the wavevector $k_x$ 
in the field components observed in Figure \ref{fig:FreeSelf.pdf}.
Similar to the analysis in Section \ref{sec:Scattering}, the increase in momentum ($p_x^2$)  can also be 
observed by considering the relative proportion of $\psi_1$ and $\psi_4$ in the overall field strength.
Figure \ref{fig:SimplePsi14.pdf} clearly shows a relative increase in $\psi_4^2$, which is expected since 
$\psi_4^2 \propto p_x^2 \propto k_x^2$.
%

%This acceleration should not be surprising when one considers that the fermion field can
%be a represention of many charged quasiparticles. 
%%
%Considering the motion of a single charged particle, 
%%
%\begin{equation}
% \frac{d^2x^\rho}{d\tau^2} =
%%+ 
%\frac{q}{m} g^{\rho\gamma}F_{\gamma\mu}\frac{dx^\mu}{d\tau},  \label{eqn:GeodesicEquation}
%\end{equation}
%%
%one obtains solutions
%%
%
%%
%\begin{eqnarray}
%t(\tau) &=& \left( \frac{m c}{q E_x}\right)  \sinh\left(  \frac{q E_x}{m c} \tau\right) \text{ and}  \\
%x(\tau) &=&  \left( \frac{m c^2}{q E_x}\right) \cosh\left(  \frac{q E_x}{m c} \tau\right),
%\end{eqnarray}
%%
%which describe Rindler-like trajectories  resulting from a constant spatial acceleration 
%%
%%
%\begin{eqnarray}
%g &=&  \frac{q E_x}{m }.
%\end{eqnarray}
%%
%The trajectories   only take this simple form because of  the high degree of symmetry 
%of the simple infinite plane charge distribution and resulting  {constant} $E_x$ outside of the pulse.  
%%
%In axisymmetry (line charge)  or spherical symmetry (point charge), 
%the electric field has a fall-off whereas here it remains constant.  
%%
%In the case described in Section \ref{sec:Scattering}, the field configuration and dynamics are not so symmetric
%and there is a dynamic mix of time-varying 
%radial and axial electromagnetic fields that accelerate the charged particles while they are confined within a large
%trap that is spatially flat (not harmonic).
%%, more akin to a rotating condensate trap \cite{Castin_RotatingTrap}.
%

Despite its simplicity, this toy model helps demonstrate how a self-interacting fermion field can evolve 
from a simple smoothly varying Gaussian distribution 
to a  distribution with high-frequency components
when in the presence of persistent strong electromagnetic fields.
{\color{blue}
Again, it is important to emphasize  that even if high-wavevector components do emerge,
such components are manageable and evolutions can be 
termintated when the conservation of energy or $\psi^\dag\psi$ deviates beyond an 
acceptable tolerance.  
}

\vspace{20mm}

%\cite{Honda_PhysRevD.102.056011}
%\cite{SHELLARD_1988262}
%\cite{Dziarmaga_PhysRevD.49.5609}
%\cite{Myers_PhysRevD.45.1355}
%\cite{Abrikosov_OrigVortex}
%\cite{Abrikosov_RevModPhys.76.975}
%\cite{deVega_ClassicalVortexSolution}
%\cite{MBHindmarsh_1995}
%\cite{RUBACK_1988669}
%\cite{MORIARTY_1988411}
%\cite{Classen_PhysRevB.93.125119}
%\cite{Wehling_02012014}
%\cite{Ye_MassiveDiracKagome}
%\cite{Yang_MassiveDirac}
%\cite{Kleidis_ChargedCosmicStrings}
%\cite{Srivastava_PhysRevD.46.1353_BUBBLES}
%\cite{Bukov_PhysRevB.89.094502}
%\cite{Albus_PhysRevA.68.023606}
%\cite{Lewenstein_PhysRevLett.92.050401}
%\cite{Krasnov_BFHModelHilbertSpace}
%\cite{Cramer_PhysRevLett.93.190405}
%\cite{Fehrmann_200423}
%\cite{Milczewski_PhysRevA.105.013317}
%\cite{Lin_PhysRevB.102.155103}
%\cite{Kleidis_ChargedCosmicStringsGravity}
%\cite{Lozano_PhysRevD.38.601}
%\cite{deVega_PhysRevD.18.2932}
%\cite{deVega_ClassicalVortexSolution}
%\cite{Jackiw_1981681_ZeroModes}
%\cite{Nielsen_197345_Original}
%\cite{Gleiser_PhysRevD.76.041701} % vav creating bubbles?
%\cite{Weigel_PhysRevLett.106.101601}
%\cite{Manton_Topological_solitons_2004tk}
%\cite{Castin_RotatingTrap}
%\cite{Lv_PhysRevA.90.034101}
%\cite{Nohl_PhysRevD.12.1840}
%\cite{Helfer_PhysRevD.99.104028}
%\cite{Abbott_PhysRevD.97.102002}
%\cite{Pillado_PhysRevD.100.023535}
%\cite{CRIBIER1964106}

\bibliography{ChargedVortexPaper}

\end{document}